\documentclass{aa}  

\usepackage{graphicx}
\usepackage{txfonts}
\usepackage{xcolor}
\usepackage{placeins}
\usepackage{cuted}
\usepackage{caption}
\usepackage{ragged2e}
\usepackage{blindtext}
\usepackage{hyperref}

\AtBeginDocument{
  \hypersetup{
    colorlinks=true,
    linkcolor=blue,
    citecolor=blue,
    urlcolor=blue
  }
}

\usepackage{graphicx}
\usepackage{float}

\begin{document} 

\title{Short-term optical monitoring of quadruply imaged quasars}    
                                                             
\author{L. J. Goicoechea\inst{1,2}
                \and
                V. N. Shalyapin\inst{2,3}
                \and
                A. Esteban-Guti\'errez\inst{4}
                \and
	            R. Gil-Merino\inst{5}
                \and 
                M. Serra-Ricart\inst{6,7,8}  
                }

\institute{Instituto de F\'isica de Cantabria (CSIC-UC), Avda. de Los                       Castros s/n, E-39005 Santander, Spain
                \and
		          Departamento de F\'\i sica Moderna, Universidad de Cantabria, 
                Avda. de Los Castros s/n, E-39005 Santander, Spain\\
                \email{goicol@unican.es}
                \and
                O.Ya. Usikov Institute for Radiophysics and Electronics, National Academy of Sciences of Ukraine, 12 Acad. Proscury St., UA-61085 Kharkiv, Ukraine
                \and
                Instituto de F\'isica y Astronom\'ia, Universidad de Valpara\'iso, Avda. Gran Breta\~na 1111, Valpara\'iso, Chile
                \and
                Escuela Superior de Ingenier\'ia y Tecnolog\'ia, Universidad Internacional de La Rioja (UNIR), Avda. Gran V\'ia Rey Juan Carlos I 41, E-26005 Logro\~{n}o, Spain\\
                \email{rodrigo.gilmerino@unir.net}
                \and
                Light Bridges S.L., Carretera del Observatorio del Teide s/n, E-38500 G\"uimar, Spain
                \and
                Instituto de Astrof\'isica de Canarias, c/ V\'ia L\'actea s/n, E-38205 La Laguna, Spain
                \and 
                Departamento de Astrofísica, Universidad de La Laguna (ULL), E-38206 La Laguna, Tenerife, Spain}
 
\abstract{
We present two case studies to demonstrate that short-term optical monitoring campaigns of quadruply imaged quasars (quads) reveal their astrophysical potential. Using the Liverpool Telescope and the third facility of the Two-meter Twin Telescope (TTT3), we monitored the quads \object{QSO 2237+0305} and \object{PS J0147+4630} in the $gr$ bands between July 2025 and January 2026. For \object{QSO 2237+0305}, the resulting light curves revealed microlensing variability over a timescale of $\sim$150 days. The $g$-band difference light curve between images A and B was then used to reliably constrain the half-light radius of the $g$-band source within the 3.5$-$16 lt-day range (1$\sigma$ confidence level). For \object{PS J0147+4630}, we derived new constraints on the AB and AC time delays from the $r$-band light curves of the ABC images in an arc-like configuration, obtaining 2$\sigma$ uncertainties of $\sim$2 days. These constraints were used to shed light on the lensing mass distribution. Overall, our results validate the design of short-term optical monitoring campaigns of quads as valuable astrophysical tools in the era of large-scale surveys. The proof of concept relies on observations with ground-based 2m-class telescopes that lack sophisticated systems to correct for optical distortions caused by atmospheric turbulence, and we indicate the astro-photometric properties of the targets and the monitoring strategies needed to achieve success with these telescopes. Additionally, this work represents the first validation of the TTT3 facility for precision photometric monitoring of gravitationally lensed quasars. 
}


\maketitle 

\section{Introduction}
\label{sec:intro}

Gravitationally lensed quasar (GLQ) research requires time-domain studies that are crucial to improve our knowledge of cosmological parameters \citep[time-delay cosmography; e.g.,][]{Birr24} and lensing mass distributions \citep[e.g.,][]{Shal25}, as well as the spatial structure of quasar accretion discs and the composition of galaxies \citep[via microlensing variability analyses; e.g.,][]{Vern24}. Additionally, several ongoing surveys of the northern sky will increase the number of known GLQs by about one order of magnitude \citep[e.g.,][]{Beni14,Duco26,Stor26}, and short-term optical monitoring campaigns of the new GLQs may be used to assess their relevance to certain astrophysical studies and to properly design their long-term follow-up.    

In order to validate the use of short-term optical monitoring to characterise the astrophysical potential of quadruply imaged quasars (quads), we conducted a six-month observational programme for two quads with very different image arrangements: \object{QSO 2237+0305} \citep{Huch85} and \object{PS J0147+4630} \citep{Berg17}, i.e., the so-called Einstein Cross and the Andromeda Parachute, respectively. The four images of the Einstein Cross (A-D; $r \sim$ 18$-$19.5 mag) are arranged in a nearly symmetric cross configuration around the bright nucleus of a spiral galaxy, whose gravity acts as a lens that deflects the light from the distant quasar. For \object{QSO 2237+0305}, the typical separation between an image and the galaxy nucleus is about 1\arcsec. Differently, the three brightest images of the Andromeda Parachute (A-C; $r \sim$ 16$-$16.5 mag) form a compact arc, and the faintest image D with $r \sim$ 18.3 mag is about 3\arcsec\ from this arc. The early-type lens galaxy of \object{PS J0147+4630} is very faint, with a magnitude of $\sim$22 in the $r$ band. 

These two GLQs were monitored in two optical bands with two 2m-class robotic telescopes from July 2025 to January 2026. In addition to the 2.0 m Liverpool Telescope \citep[LT;][]{Stee04} at Roque de los Muchachos Observatory, which was a key facility of the Gravitational LENses and DArk MAtter project\footnote{\url{https://gravlens.unican.es}} \citep[GLENDAMA;][]{Gilm18}, we used the 2.0 m TTT3 telescope. TTT3 had its first light in 2025 and is producing science during the commissioning phase \citep[e.g.,][]{Fuen25}. This facility is part of the Two-meter Twin Telescope (TTT)\footnote{\url{https://ttt.iac.es}} array at Teide Observatory. 

The structure of the paper is as follows. Our LT+TTT3 observations and light curves of \object{QSO 2237+0305} are described in Sect.~\ref{sec:cross}. In Sect.~\ref{sec:crossmicro} we discuss the variability of the Einstein Cross and reliably estimate the size of its $g$-band source from a microlensing variability analysis. In Sect.~\ref{sec:parach} we present our LT+TTT3 observations and light curves of \object{PS J0147+4630}. Based on the short-term variability of the Andromeda Parachute, Sect.~\ref{sec:parachvar} includes new constraints on the time delays between arc images, and a discussion of the mass distribution in the lensing galaxy and along the line of sight to the system. Finally, in Sect.~\ref{sec:end} we present our conclusions.

\section{Optical monitoring of \object{QSO 2237+0305}}
\label{sec:cross}

\subsection{Observations}

In the framework of the GLENDAMA and the Gravitational Lenses And Microlensing (GLAM; PI: R. Gil-Merino) projects, we observed \object{QSO 2237+0305} with LT/IO:O over the period from 18 July to 24 October 2025, obtaining a 300 s frame in the $g$ band and another of 180 s in the $r$ band on each observing night. The optical wide field camera IO:O provides a field of view of $10\arcmin \times 10\arcmin$ with a pixel scale of 0\farcs30 (binning $2 \times 2$).

\begin{figure}
\centering
\includegraphics[width=9cm]{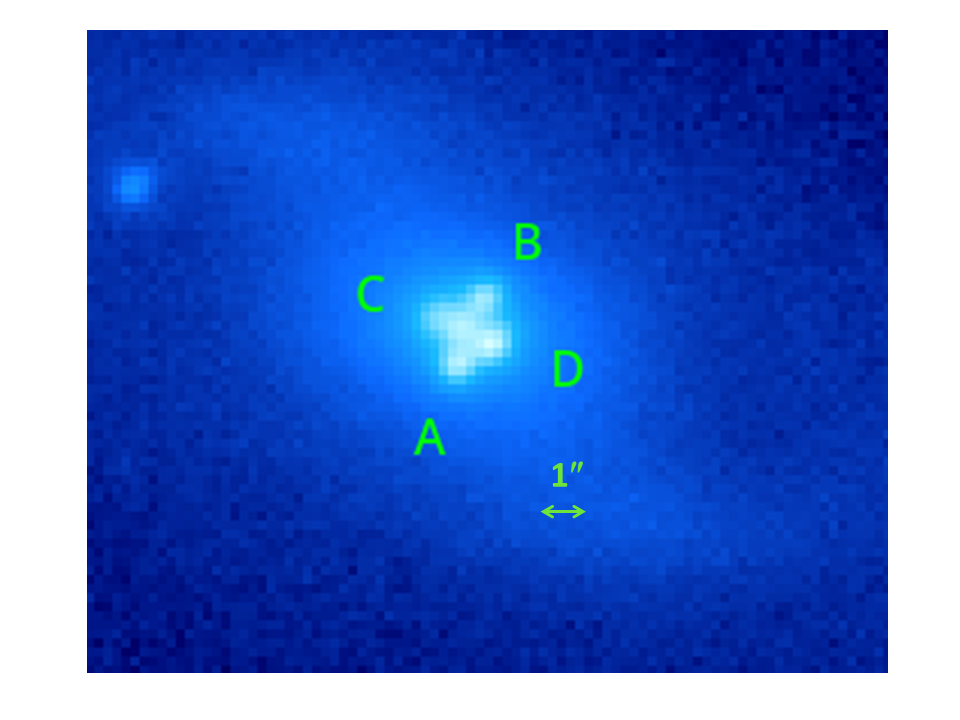}
\includegraphics[width=9cm]{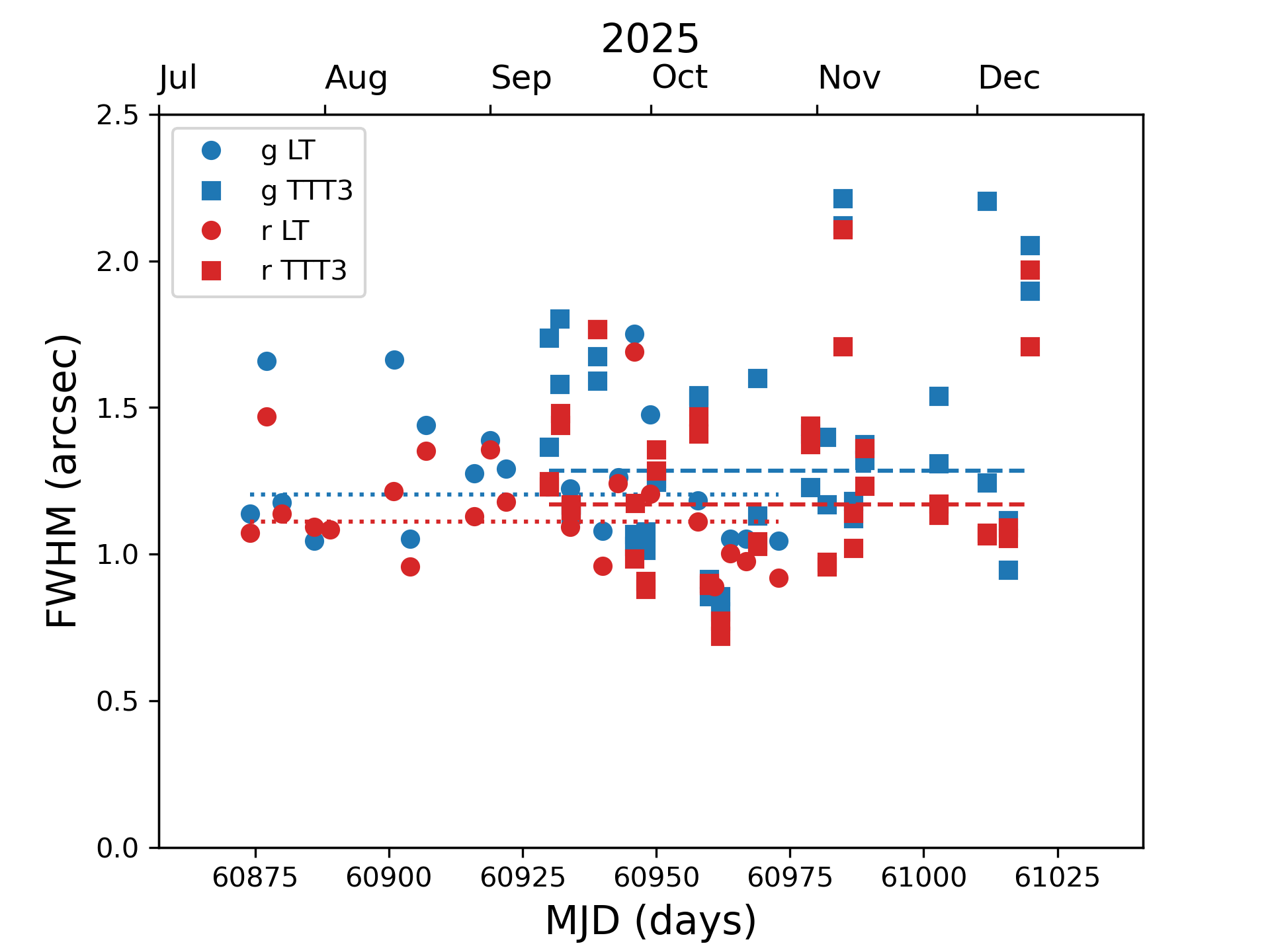}
\caption{Imaging of \object{QSO 2237+0305}. Top panel: the Einstein Cross (ABCD) from a TTT3 $r$-band frame taken on 13 October 2025 with FWHM = 0\farcs77. This FWHM is less than the angular separation between images A and D ($\sim$1\arcsec). Bottom panel: comparison of FWHM distributions using LT and TTT3 observations. The dotted and dashed horizontal lines show the LT and TTT3 median values, respectively.}
\label{fig:cross_ima_see}
\end{figure}

The TTT3 telescope is equipped with the CCD Observatory for Light with Optical Robotic Six-band photometry (COLORS), which is a CCD camera with a field of view of 7\farcm85 $\times$ 7\farcm85 and a 0\farcs23~pixel$^{-1}$ scale. As part of the performance testing programme GRAVLENS25 (PI: L.J. Goicoechea), the TTT3/COLORS observations of \object{QSO 2237+0305} were made from 11 September to 10 December 2025, so there is a month and a half overlap between the LT and TTT3 monitoring campaigns. Each observing night, two 300 s exposures were taken with the Sloan $g$ filter and two 180 s exposures with the $r$ filter. In the top panel of Figure~\ref{fig:cross_ima_see}, we show the four quasar images A-D around the nucleus of the spiral lens galaxy from a TTT3 $r$-band frame taken under sub-arcsecond seeing conditions. A full-width at half maximum (FWHM) of the seeing disc less than 1\arcsec\ allows us to see the four images and the galaxy nucleus separately. 

The FWHM distributions from both telescopes at two different observatories were very similar (see the bottom panel of Figure~\ref{fig:cross_ima_see}), with median values of 1\farcs26 (LT) and 1\farcs37 (TTT3) in the $g$ band, and 1\farcs15 (LT) and 1\farcs24 (TTT3) in the $r$ band. For each optical band, a difference of $\sim$0\farcs1 is much smaller than the measured medians. We also note that the TTT array consists of two identical 2.0-m  telescopes (TTT3 and TTT4) and two smaller replicas of 0.80-m (TTT1 and TTT2; in operation since 2023). These 0.80-m telescopes are being used to study the variability of GLQs with optically bright and widely separated images. For example, \citet{Osco96} and \citet{SerraRicart99} achieved 2\% photometry for \object{QSO 0957+561} using the 0.8 m IAC80 Telescope. 

\subsection{Light curves}
\label{sec:crossphot}

To obtain the image light curves for each band and telescope, we need to model the observed light distributions (see the top panel of Figure~\ref{fig:cross_ima_see}) using the corresponding point spread functions (PSFs). Our photometric model is detailed in \citet{Shal26}, but we provide a brief summary here for completeness. For each sub-frame containing the four quasar images, the nearly face-on barred spiral galaxy and a background estimation region, we considered a comprehensive model including four empirical PSFs for the quasar images, the sum of a De Vaucouleurs distribution and an exponential distribution (both convolved with the empirical PSF) for the analytical structure of the galaxy, the numerical (non-analytical) galaxy structure, and a constant term for the sky background. Using this photometric scheme and the IMFITFITS software \citep{McLe98}, we extracted the A-D image fluxes in each usable frame and obtained their magnitudes on each observing night. 

\begin{figure*}
\centering
\includegraphics[width=9cm]{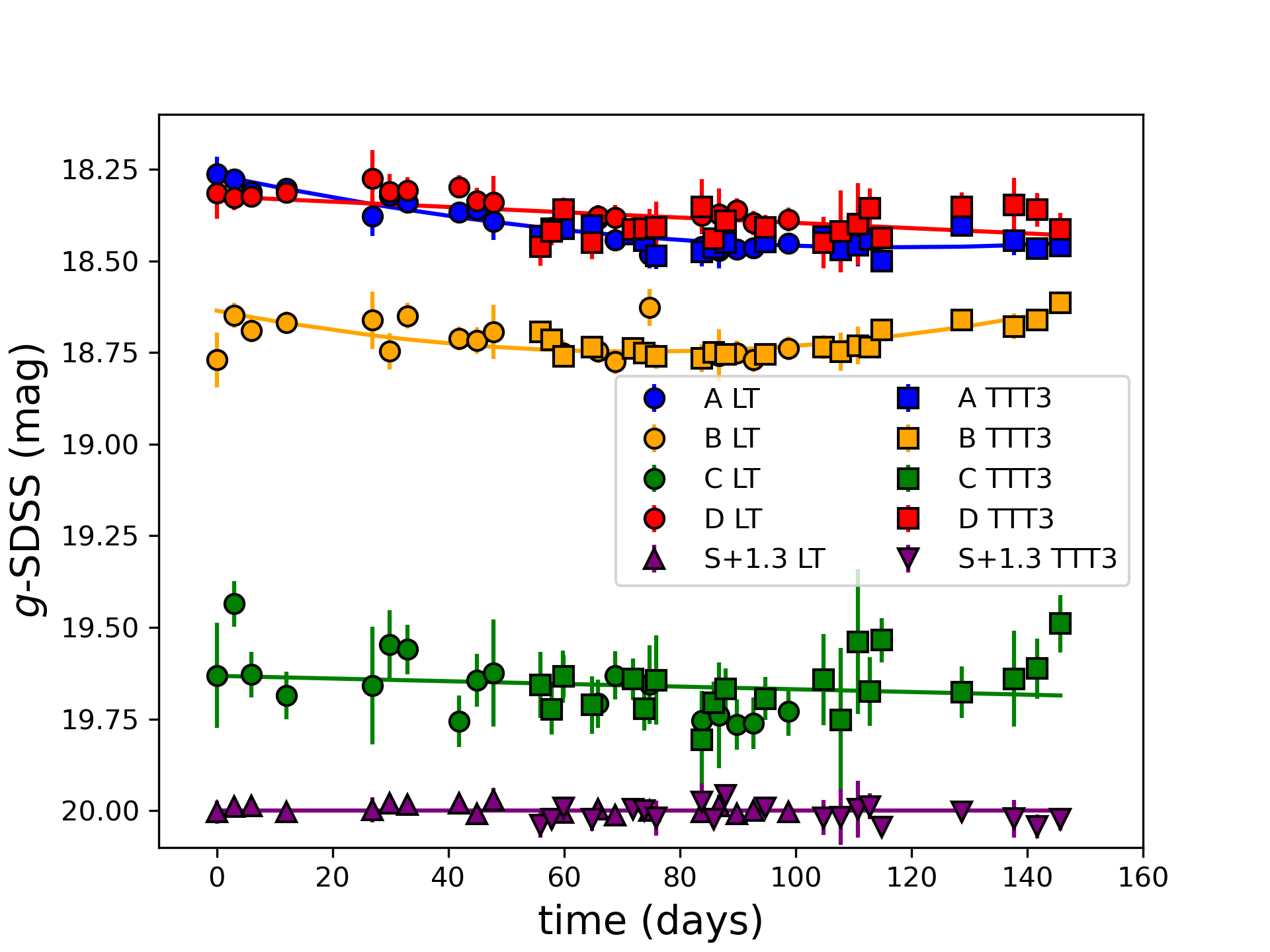}
\includegraphics[width=9cm]{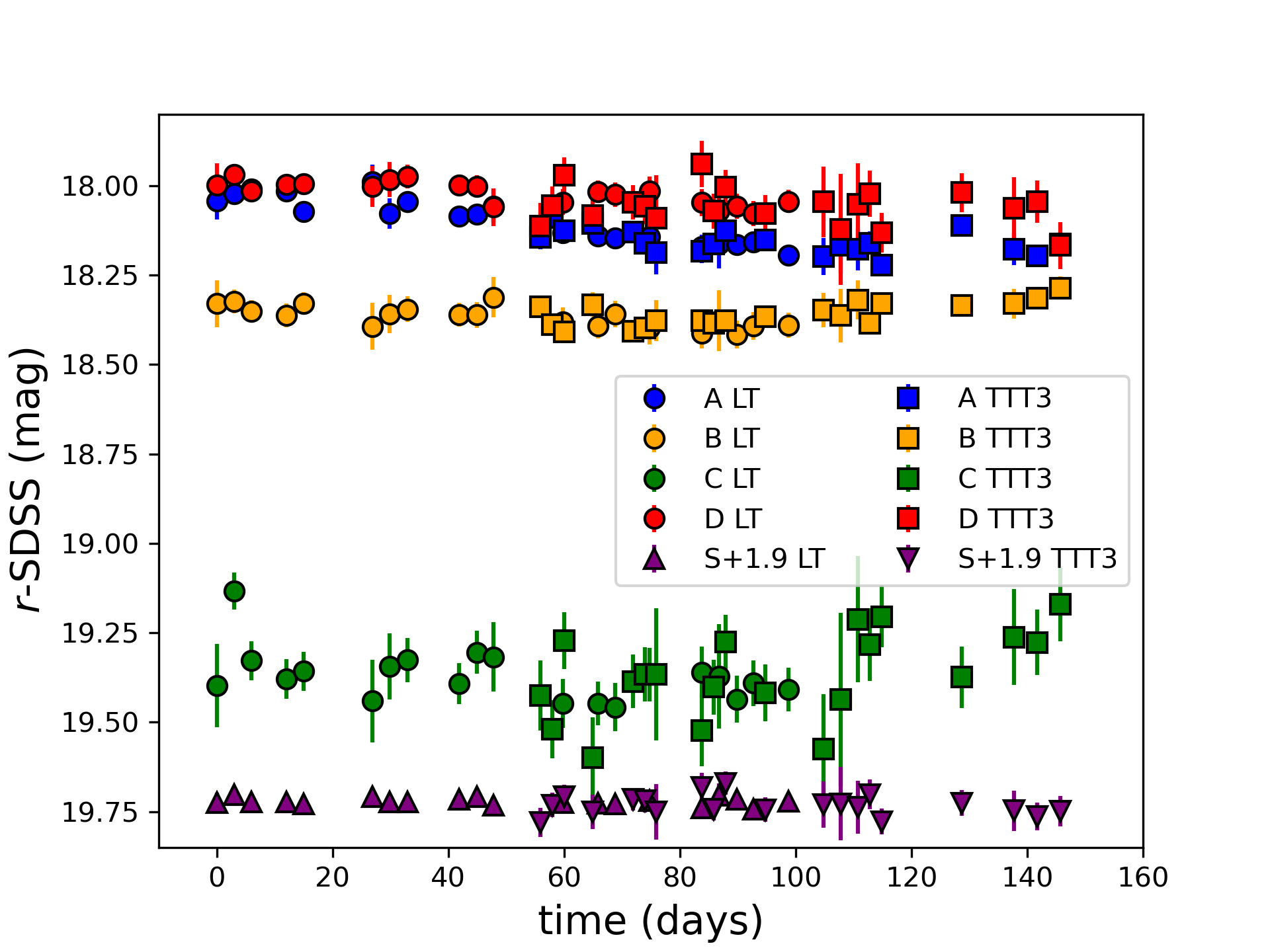}
\caption{LT+TTT3 light curves of \object{QSO 2237+0305}. The brightness records of a field star S are also shown for comparison, and $t$ = 0 corresponds to the first observing epoch on 18 July 2025. Left panel: $g$-band magnitudes. The measured magnitudes are fitted to low-degree polynomials (solid lines) as a guide to the eye. The light curve of S is shifted by +1.3 mag for clarity. Right panel: $r$-band magnitudes. The record of S is vertically shifted by +1.9 mag.} 
\label{fig:cross_lcs}
\end{figure*}

Typical magnitude errors of A-D and field star S were also estimated from standard deviations between magnitudes with time separations $\leq$2.5 d, which
were divided by the square root of 2. These typical uncertainties for each band and telescope\footnote{The LT typical errors are based on all available observations from the GLENDAMA project during the 2013$-$2025 period.} were then weighted by the inverse of the signal-to-noise ratio at every epoch to derive individual errors \citep[e.g.,][]{Howe06}. The photometric accuracy is detailed in the first two rows from the top in Figure~\ref{fig:quads_err} (see Appendix~\ref{sec:photaccu}). In the $g$ band, we achieve similar accuracy with both telescopes. It is also worth noting that the TTT3 $g$-band observations of the Einstein Cross yield a low flux uncertainty of $\sim$2.5\% for the images A and B. From the LT $r$-band frames, we achieve 2.5$-$5\% photometry for the three brightest images (A, B, and D), while the TTT3 $r$-band frames lead to a flux uncertainty of 5$-$10\% for the D image. Using TTT3 $r$-band observations, the flux of the faintest image C is also measured with a relatively high uncertainty of $\sim$10\%. As a final data product in each optical band, we obtained the LT+TTT3 combined light curves of \object{QSO 2237+0305} and S. These final records are displayed in Figure~\ref{fig:cross_lcs}, and included in Tables 1 ($g$ band) and 2 ($r$ band) at the CDS. Additionally, the inter-telescope quasar calibration is described in Appendix~\ref{sec:intertelcal}. 

\section{\object{QSO 2237+0305}: astrophysics from microlensing variability}
\label{sec:crossmicro}

Since the inter-image time delays in \object{QSO 2237+0305} are of order hours \citep[e.g.,][]{Dai03,Berd17,Fedo26}, the sampling rate used in Figure~\ref{fig:cross_lcs} does not allow us to probe these very short lags between quasar images. In addition, the polynomial fitting in the left panel of Figure~\ref{fig:cross_lcs} describes non-parallel variations of A-D that are not solely due to the intrinsic quasar signal. Although we also conducted an intra-night monitoring of \object{QSO 2237+0305} with TTT3 on 10 and 12 July 2025, unfortunately, moonlight severely contaminated the quasar images, preventing a clear identification of possible intrinsic events usable for measuring or refining the current time delays. We also want to emphasise that to measure precise delays in the Einstein Cross, X-ray data appear more appropriate than intra-night optical data \citep[e.g.,][]{Fedo26}.

\begin{figure}[h!]
\centering
\includegraphics[width=9cm]{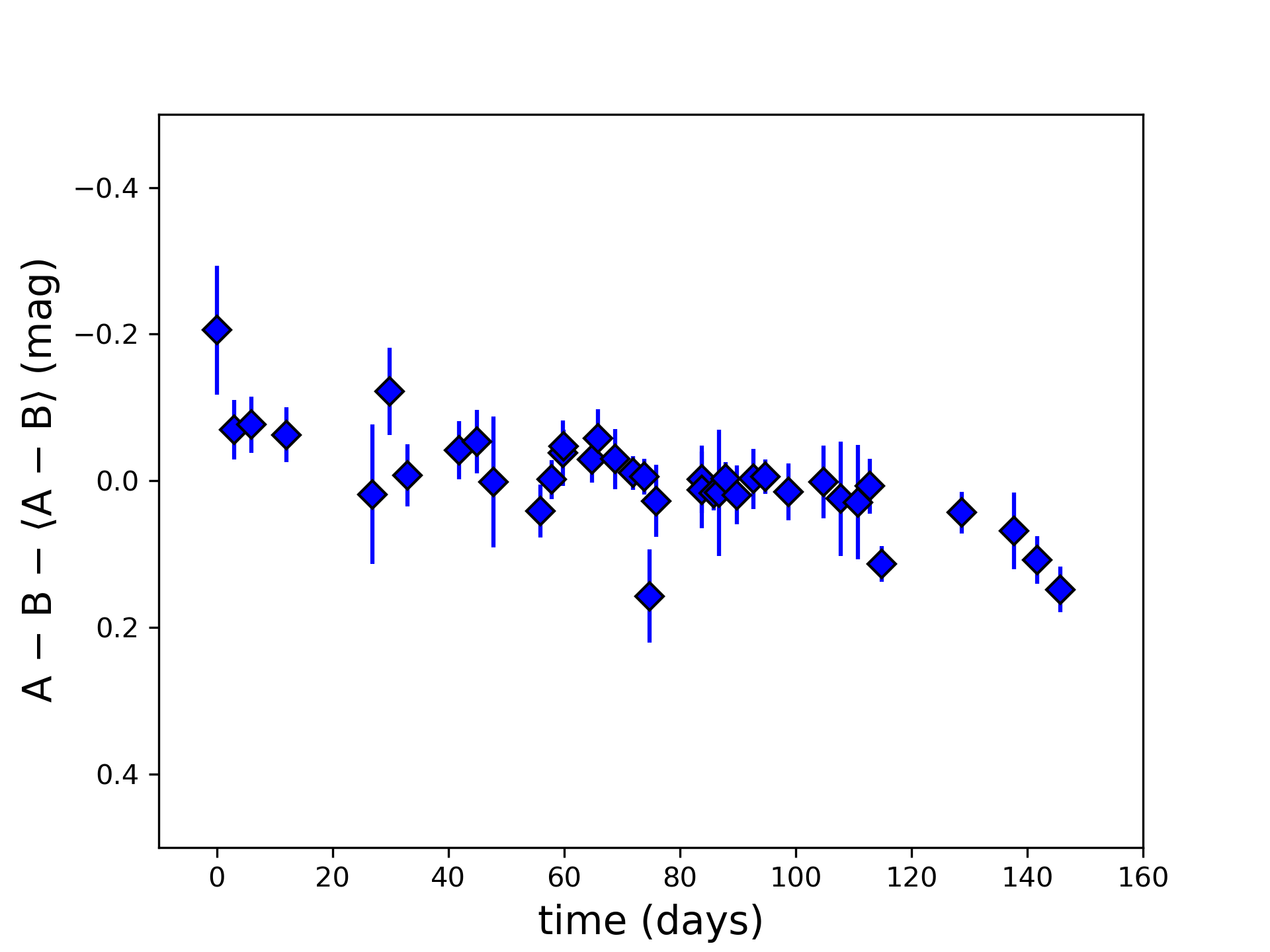}
\includegraphics[width=9cm]{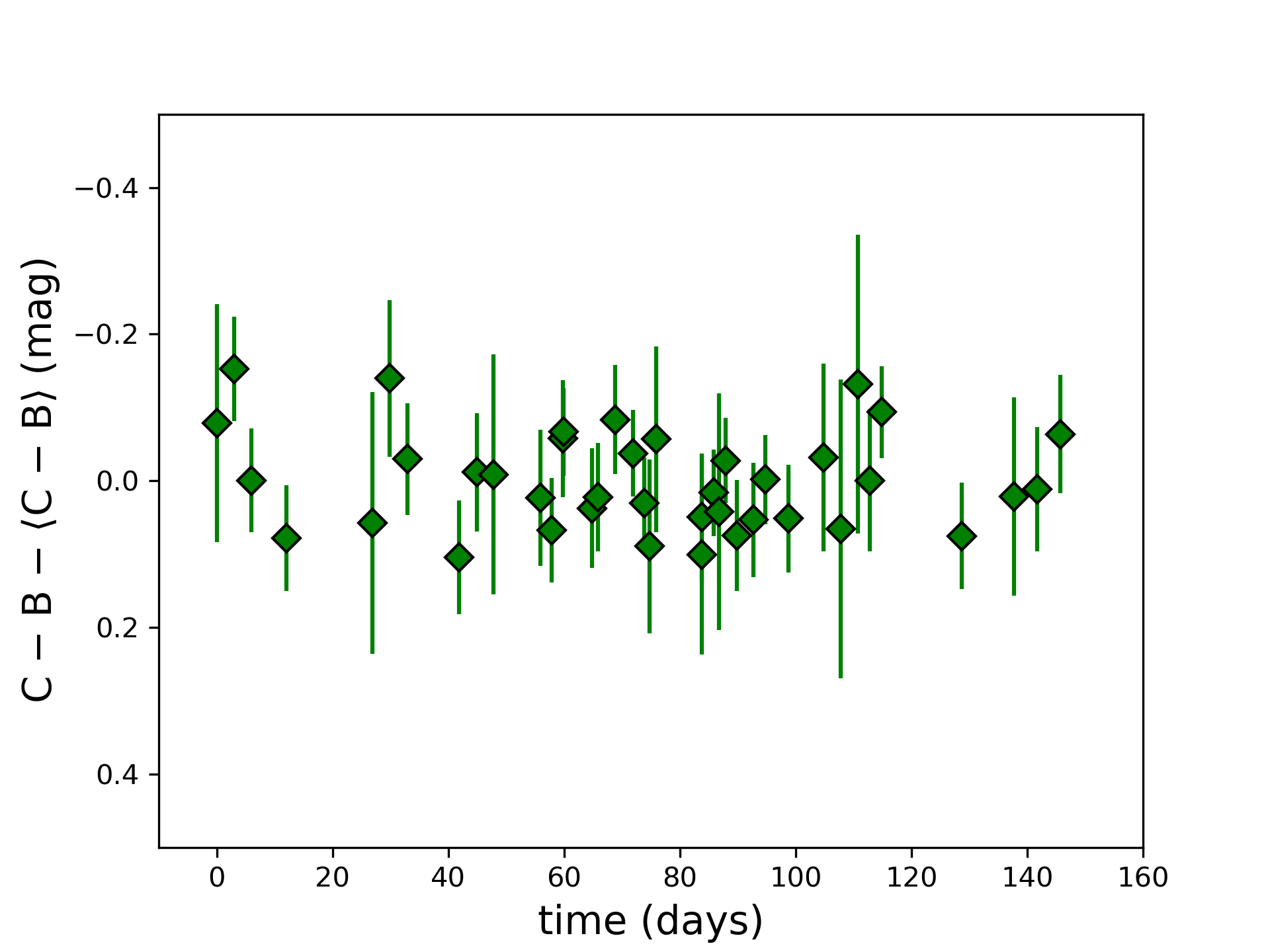}
\caption{Difference light curves of \object{QSO 2237+0305} from LT+TTT3 $g$-band observations.}
\label{fig:cross_dlcs}
\end{figure}

Our LT+TTT3 light curves in Figure~\ref{fig:cross_lcs} are thus mainly relevant for microlensing variability studies, and we focused on the $g$-band records because microlensing-induced variations are expected to be stronger at shorter wavelengths \citep[smaller sources; e.g.,][]{Vern24}. Taking the extremely short inter-image delays into account (see above), the microlensing signals can be derived from differences between image pairs, e.g., building the difference light curves (DLCs) A $-$ B $-$ $\langle$A $-$ B$\rangle$, C $-$ B $-$ $\langle$C $-$ B$\rangle$, and D $-$ B $-$ $\langle$D $-$ B$\rangle$. These DLCs remove the intrinsic variability of the quasar and the differential dust extinction \citep[e.g.,][]{Isla25}. The most accurate DLC is shown in the top panel of Figure~\ref{fig:cross_dlcs}, the least in the bottom panel. 

\begin{figure}
\centering
\includegraphics[width=9cm]{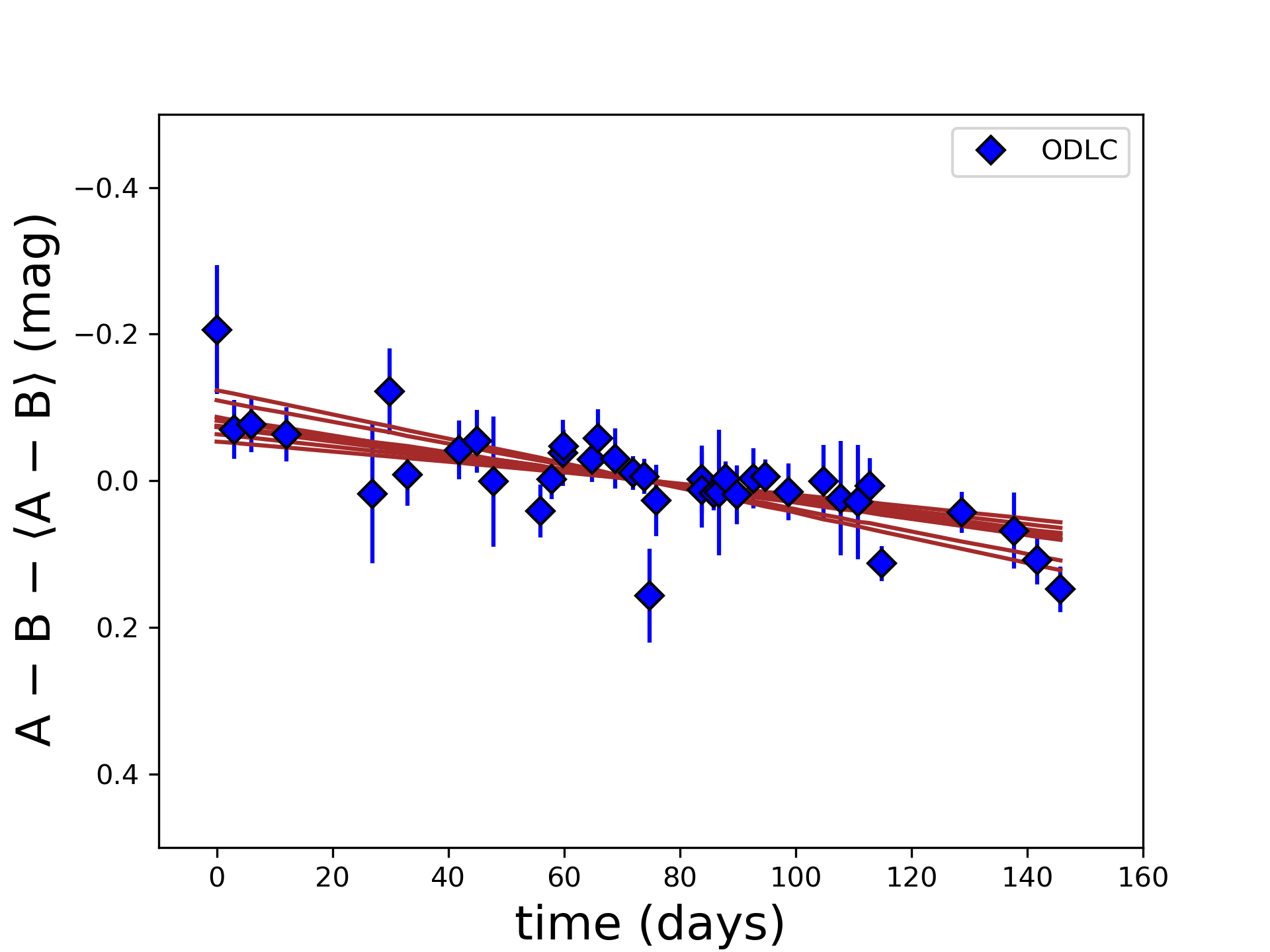}
\caption{Comparison between the observed DLC in the top panel of Figure~\ref{fig:cross_dlcs} (labelled with ODLC) and a set of synthetic DLCs yielding 0.9 $\leq$ RMS < 1.1 (brown lines). The simulated DLCs are outputs generated by the second analysis, using $x$ = 0.2 and $v_{\rm e}$ = 450 km s$^{-1}$.}
\label{fig:cross_gdlcABsim}
\end{figure}

The observed microlensing variability over the monitoring period of $\sim$150 d can be used to obtain relevant astrophysical information \citep[e.g.,][]{Gilm05}. In this paper, to constrain the size of the $g$-band source at $z_{\rm s}$ = 1.695 \citep{Huch85}, synthetic DLCs from random source trajectories on simulated magnification maps of A and B are compared with the microlensing signal in the top panel of Figure~\ref{fig:cross_dlcs}. We generated the point-source magnification maps by using the convergences and shear strengths from Table 1 of \citet{Poin10a}. We also reasonably assumed that the convergences in the bulge of the lensing galaxy at $z_{\rm l}$ = 0.039 \citep{Huch85} are exclusively due to stars, which are randomly distributed from a mass function $dN/dM \propto M^{-1.3}$ with a maximum-to-minimum mass ratio of 50 \citep{Goul00}. 

Each point-source (one-pixel size source) magnification map in the source plane with 500 pix/$R_{\rm E}$ resolution ($R_{\rm E}$ is the Einstein radius of a star with the mean mass $\langle M \rangle$) was created using the Poisson and Inverse Polygon technique\footnote{https://microlensing.overfitting.es/} \citep{Shal21}, and then convolved with Gaussian kernels $\exp(-r^2/2R^2)$, where $R$ is given in units of $R_{\rm E}$ and $R_{\rm s}$ = 1.177 $R$ is the half-light radius of the Gaussian source with typical radius $R$. Since previous studies indicated that $R_{\rm s} << R_{\rm E}$ for optical sources \citep[e.g.,][]{Koch04,Angu08,Poin10b}, in addition to a point source ($R \sim$ 0), we initially used a uniform distribution of five values of the scaled radius $x = R/R_{\rm E}$, ranging from $x$ = 0.1 to 0.5.

\begin{figure}[h!]
\centering
\includegraphics[width=9cm]{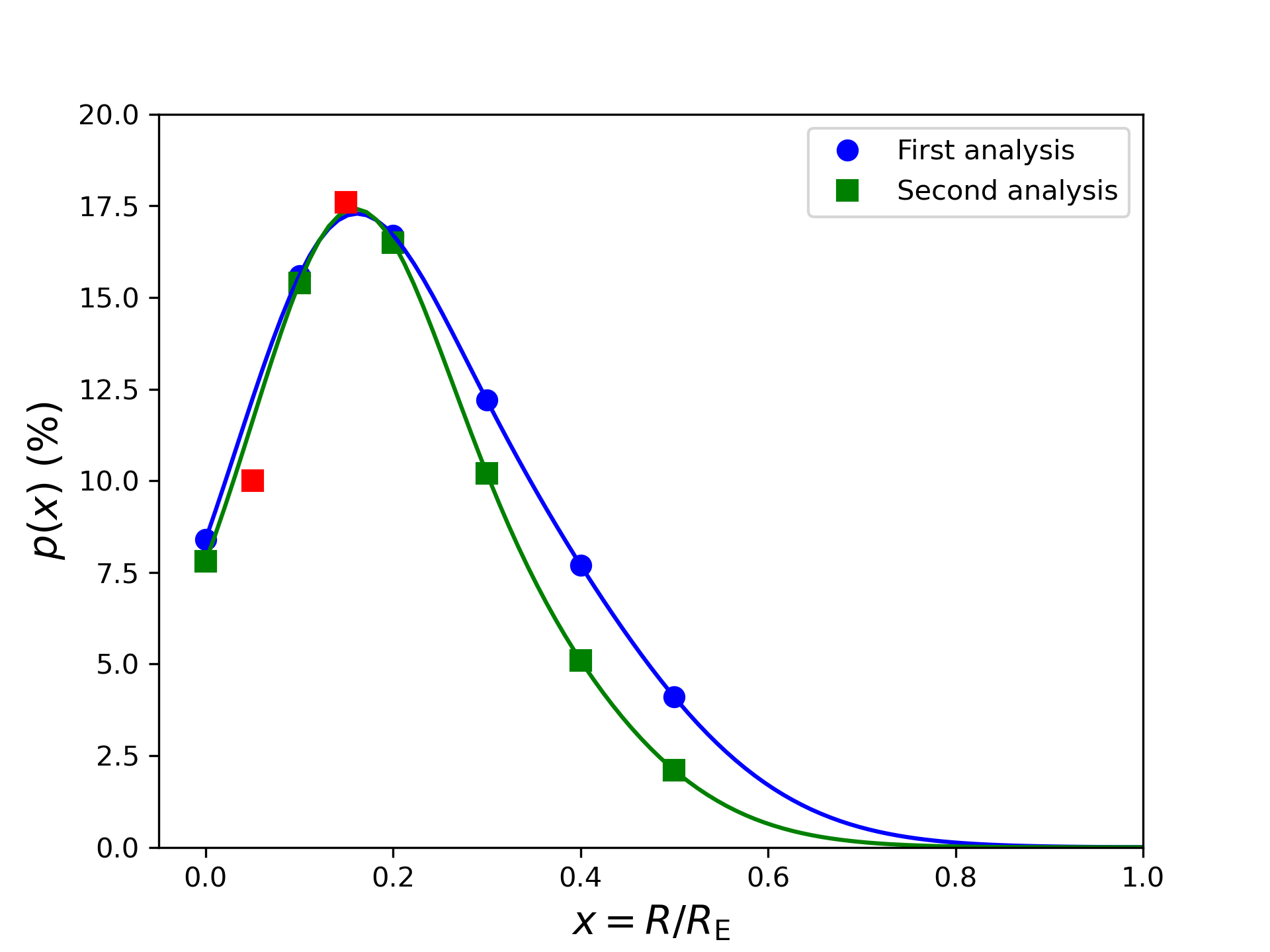}
\caption{Consistency probability distributions for $v_{\rm e}$ = 450 km s$^{-1}$. The solid lines represent fits to the results using the initial grid in $x$ and the two red squares are used to test the fit to the results of the second analysis (see the main text).}
\label{fig:cross_gdlcABcpd_cen}
\end{figure}

We also adopted a concordance flat cosmology and a mean stellar mass $\langle M \rangle$ = 0.3 $\rm{M_{\odot}}$ ($R_{\rm E}$ = 10$^{17}$ cm). Regarding the transverse velocity of the lensing galaxy $v_{\rm t}$, \citet{Poin10a} reported a probability peak around 400 km s$^{-1}$, and 68\% confidence limits $v_{\rm t} >$ 338 km s$^{-1}$ and $v_{\rm t} <$ 486 km s$^{-1}$. Therefore, we plausibly assumed that the actual value of $v_{\rm t}$ is within the range 300$-$500 km s$^{-1}$. Since our microlensing simulations do not account for random motions of stars, $v_{\rm t}$ must be increased by $\sim$50 km s$^{-1}$ \citep[e.g.,][]{Gilm05,Angu08}, and we used five effective velocities ($v_{\rm e}$ = 350, 400, 450, 500, and 550 km s$^{-1}$). 

For each source trajectory on a pair of AB maps, we obtained one synthetic difference curve A $-$ B $-$ $\langle$A $-$ B$\rangle$, which was compared with the observed signal. Only the simulated DLCs leading to a root mean square (RMS) of relative residuals of about 1 were considered to be consistent with the observations \citep[e.g.,][]{Isla25}. Thus, we generated $5 \times 10^5$ simulated DLCs for each pair of AB maps (source size) and effective velocity, and adopted the consistency interval 0.9 $\leq$ RMS < 1.1. The minimum RMS value we obtain is 0.9 and we consider a symmetric interval around RMS = 1; that is, a maximum deviation of 10\% from 1. For each value of $x$ and $v_{\rm e}$, the consistency probability was defined as $p = N_{\rm con}/N_{\rm sim}$, where $N_{\rm con}$ is the number of synthetic DLCs yielding 0.9 $\leq$ RMS < 1.1 and $N_{\rm sim}$ = $5 \times 10^5$.

We took the central value of the effective velocity range ($v_{\rm e}$ = 450 km s$^{-1}$) to check the influence of map size and edge treatment on the results. In a first analysis, we built square magnification maps of 7 $R_{\rm E}$ on a side. We also considered zero magnification outside map boundaries during convolution, and usable central squares with sides of 5 $R_{\rm E}$ ($x \leq$ 0.3) and 4 $R_{\rm E}$ (0.4 $\leq x \leq$ 0.5). In a second analysis, we created square maps with a side of 10 $R_{\rm E}$ (see Figure~\ref{fig:cross_mmaps2} in Appendix~\ref{sec:magmaps}). To better prevent biases at the edges of each convolved map, during convolution, we set the magnification outside the map to the average value inside it. Moreover, only unbiased central squares of 6 $R_{\rm E}$ on a side were used to produce random source trajectories and simulated DLCs. 

\begin{figure}[h!]
\centering
\includegraphics[width=9cm]{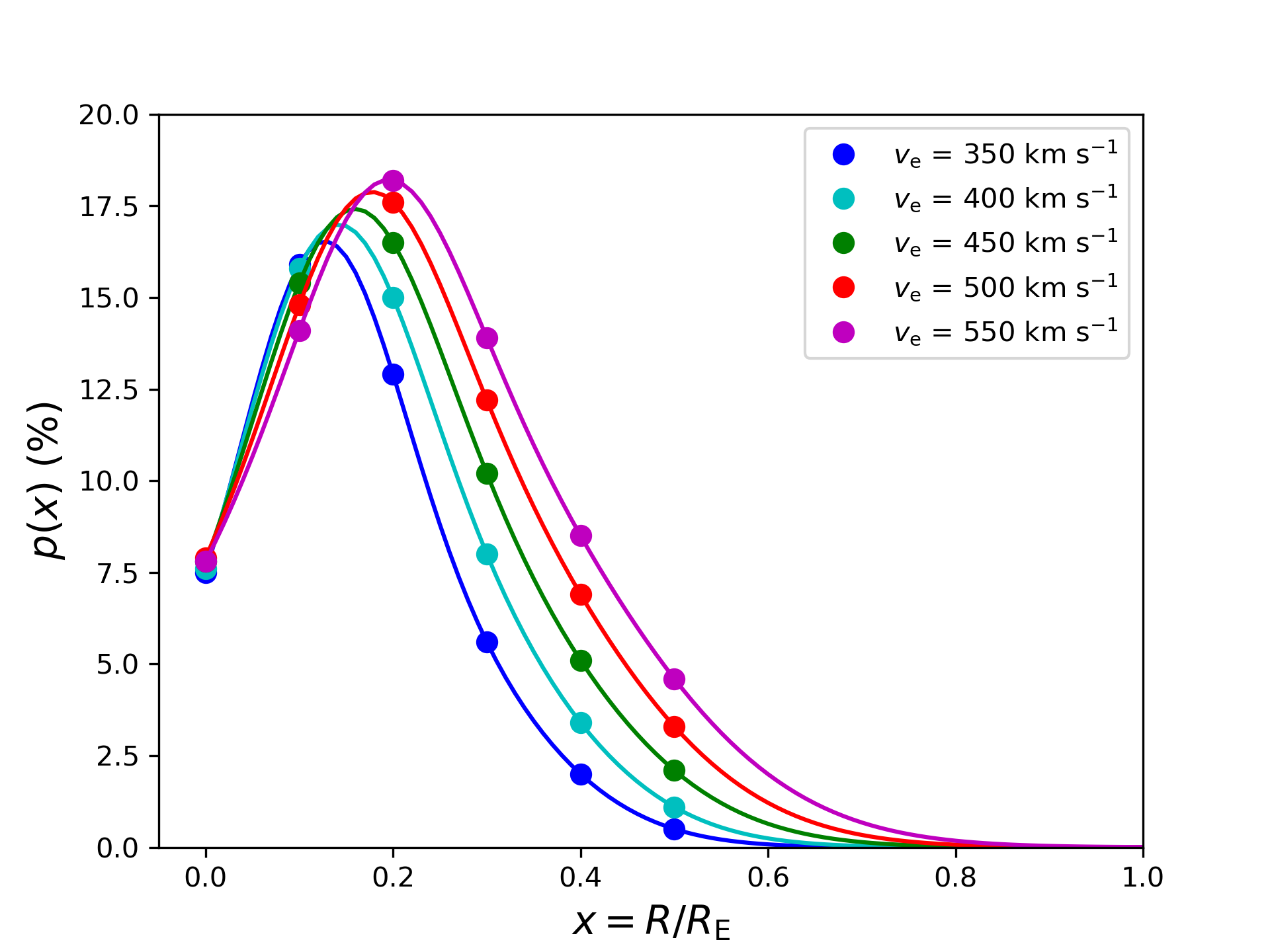}
\includegraphics[width=9cm]{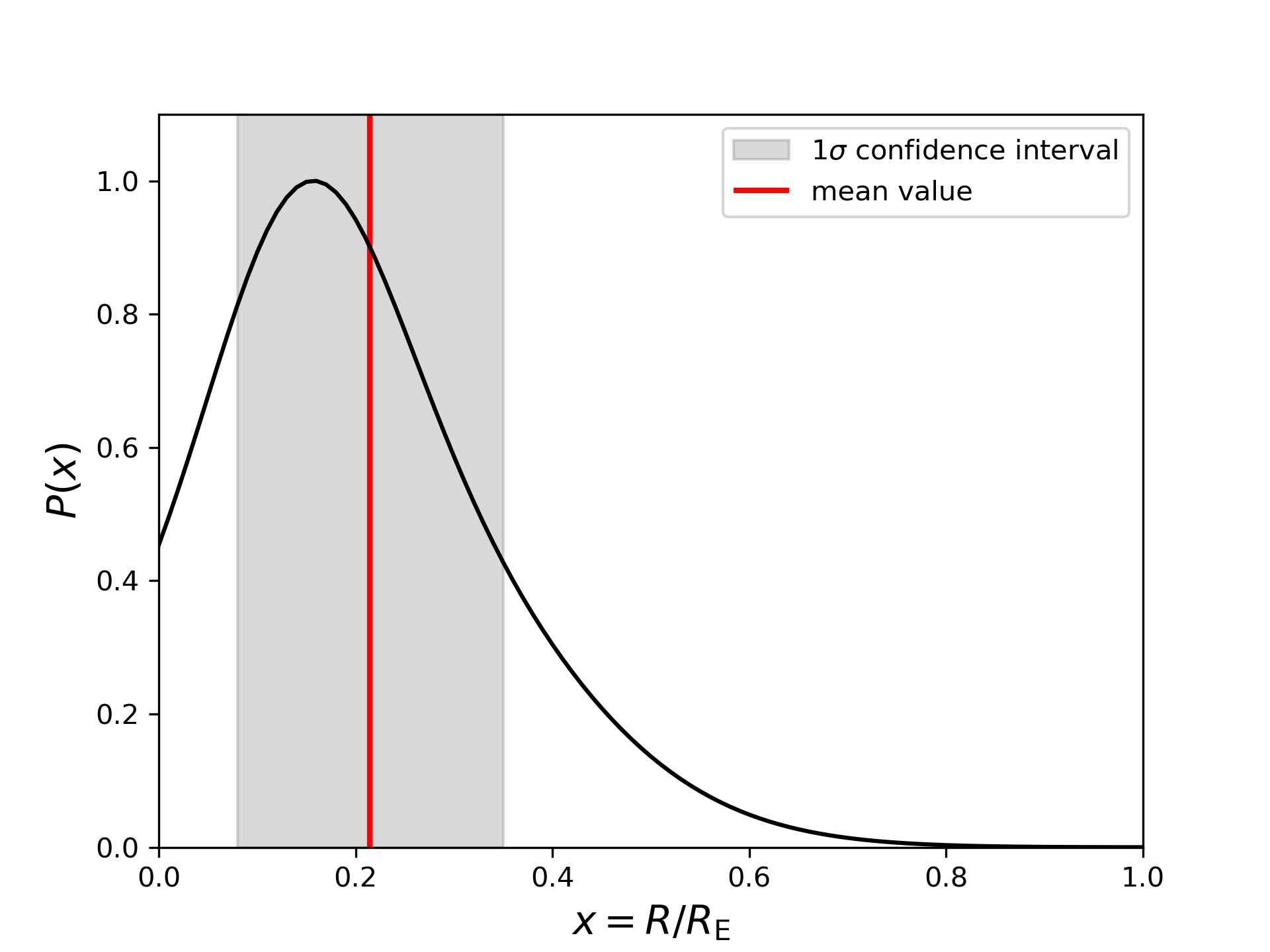}
\caption{Probability distributions for the scaled radius of the $g$-band source in \object{QSO 2237+0305}. Top panel: consistency probability distributions for effective velocities within the range 350$-$550 km s$^{-1}$. Bottom panel: probability distribution by marginalising over $v_{\rm e}$ (arbitrary units).}
\label{fig:cross_gdlcABcpd_all}
\end{figure}

Figure~\ref{fig:cross_gdlcABsim} shows randomly chosen synthetic DLCs for $x$ = 0.2 and $v_{\rm e}$ = 450 km s$^{-1}$ that fit the observations well. Although this set of simulated DLCs corresponds to the second analysis, the first analysis yields very similar results. For $v_{\rm e}$ = 450 km s$^{-1}$, the consistency probability distributions $p(x)$ are also displayed in Figure~\ref{fig:cross_gdlcABcpd_cen}. We fitted a two-Gaussian mixture model to the two discrete distributions for our initial grid of $x$ values (blue and green lines). Additionally, in the second analysis, we considered two extra scaled radii of 0.05 and 0.15 (red squares) to discuss how the fitting (interpolation) procedure works. The interpolated distribution is slightly broader than the discrete one (eight squares), which should lead to an overestimation of confidence intervals for the source size. 

We note that the second analysis is based on larger maps and better edge treatment, and it produces a narrower discrete distribution. In addition, using several samples of $5 \times 10^5$ simulated DLCs, we verified that the uncertainty in $p(x)$ is less than 0.1\%. Thus, we focused on the methodology of the second analysis to obtain the final discrete and interpolated distributions for all values of $v_{\rm e}$ (see the top panel of Figure~\ref{fig:cross_gdlcABcpd_all}). The bottom panel of Figure~\ref{fig:cross_gdlcABcpd_all} also shows the marginal probability distribution $P(x) \propto \int p(x \,|\, v_{\rm e}) \, dv_{\rm e}$. Using this last distribution, the 1$\sigma$ confidence interval for the half-light radius of the $g$-band source is $R_{\rm s}$ = 0.25 $\pm$ 0.16 $R_{\rm E}$, in very good agreement with a recent measurement from a long-term monitoring in the $g$ band \citep{Shal26}.

\section{Optical monitoring of \object{PS J0147+4630}}
\label{sec:parach}

\subsection{Observations}

\begin{figure*}[!t]
\sidecaption
\includegraphics[width=5.7cm]{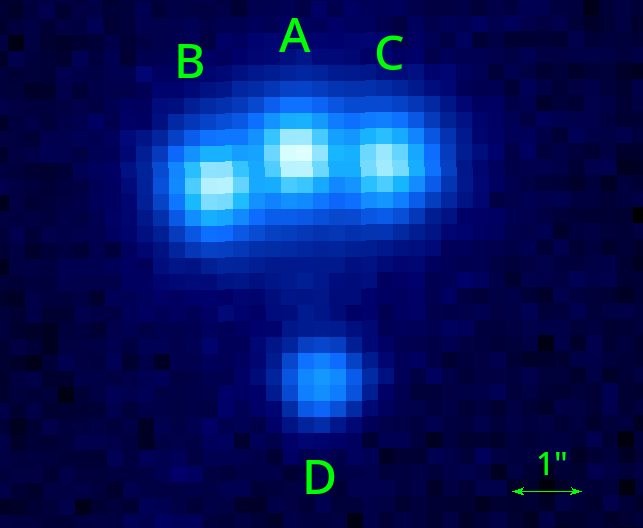} 
\includegraphics[width=6.3cm]{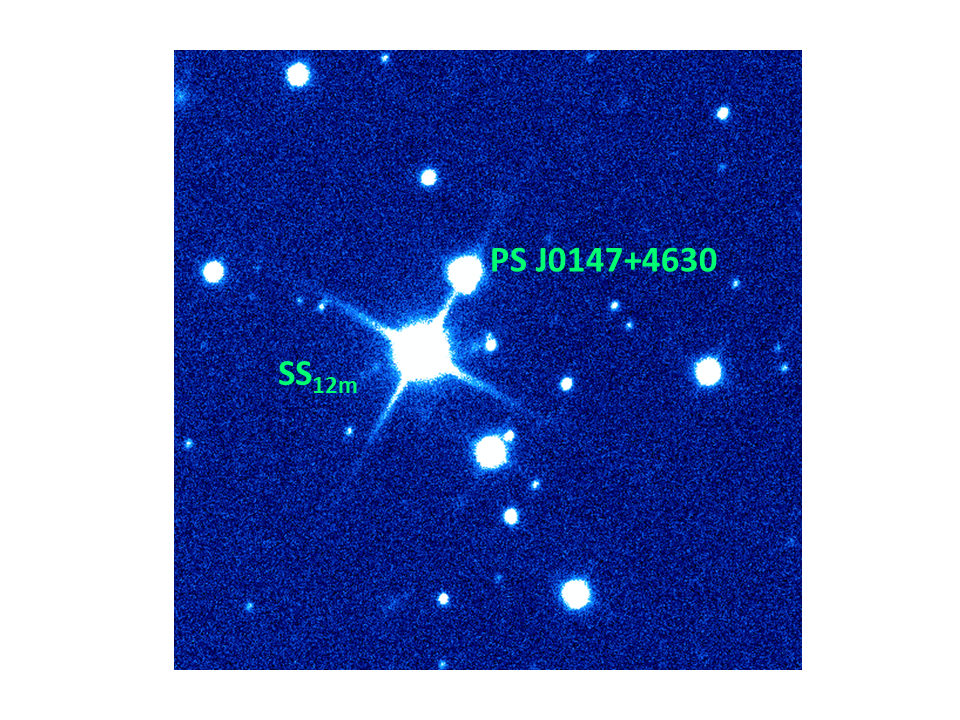}
\caption{TTT3 $r$-band imaging of \object{PS J0147+4630}. Left panel: the Andromeda Parachute (ABCD) from a TTT3 frame taken on 19 October 2025 with FWHM = 0\farcs65. Right panel: spike of the saturated star SS$_{\rm 12m}$ contaminating the lens system in a frame taken on 11 January 2026.}
\label{fig:parach_ima_spike}
\end{figure*}

\begin{figure*}
\centering
\includegraphics[width=8cm]{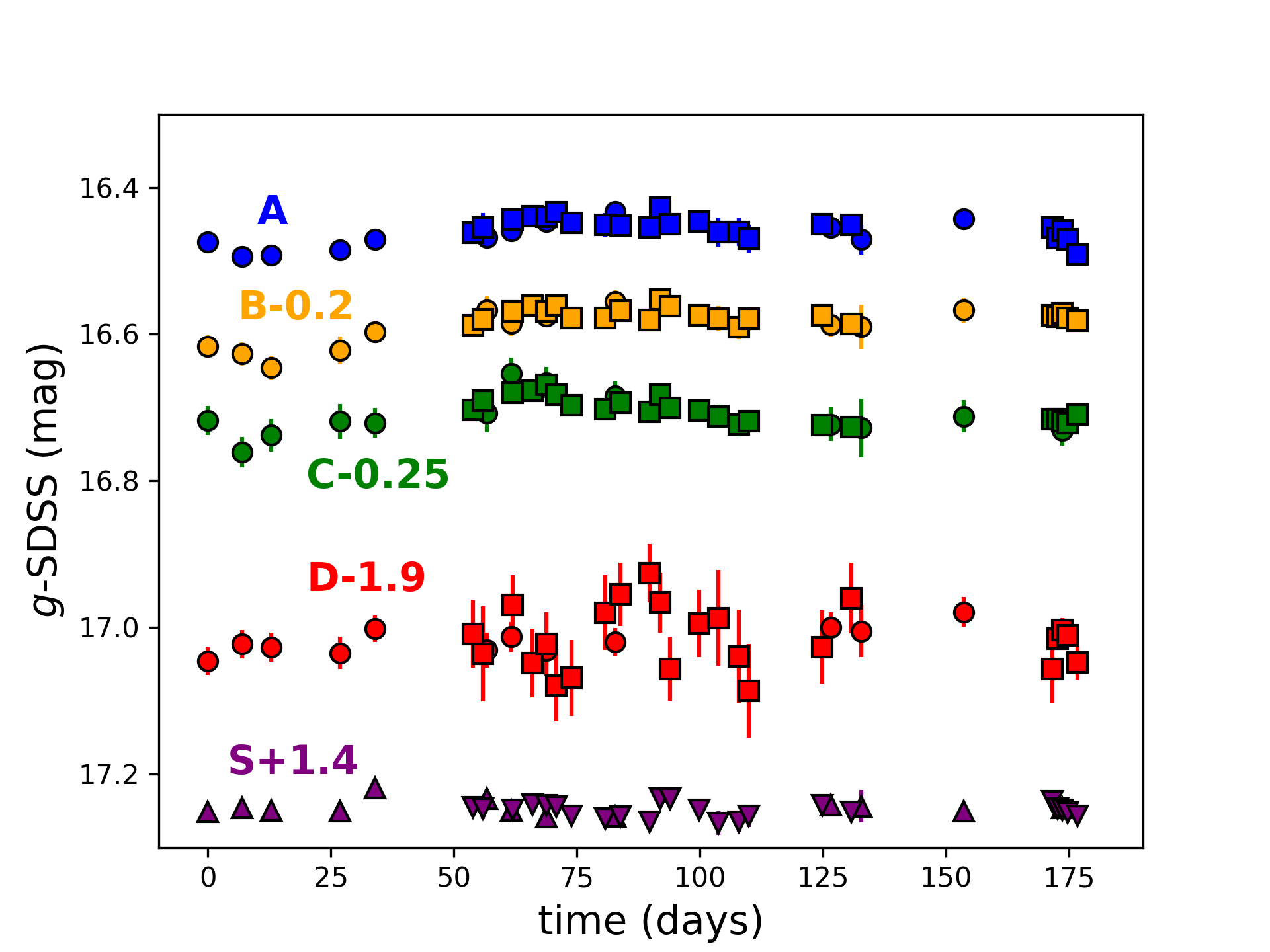}
\includegraphics[width=8cm]{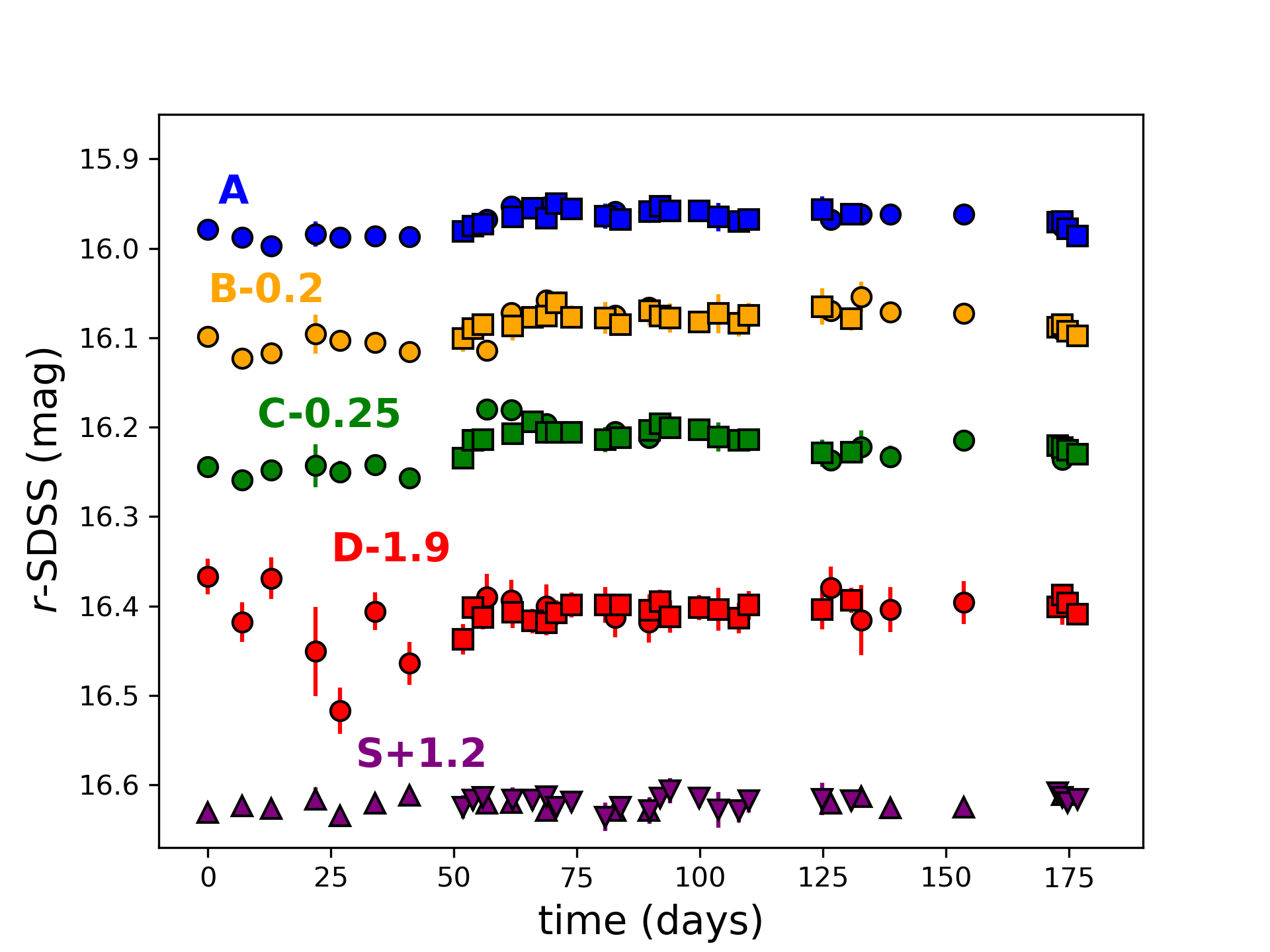}
\caption{Light curves of \object{PS J0147+4630} and a field star S. The circles and up-pointing triangles correspond to LT observations, whereas the squares and down-pointing triangles are TTT3 data. Here, $t$ = 0 represents the first observing epoch on 22 July 2025. Left panel: $g$-band magnitudes. The measured magnitudes are shifted by $-$0.2 (B), $-$0.25 (C), $-$1.9 (D), and $+$1.4 (S). Right panel: $r$-band magnitudes. The brightness records are shifted by $-$0.2 (B), $-$0.25 (C), $-$1.9 (D), and $+$1.2 (S) mag.}
\label{fig:parach_lcs}
\end{figure*}

As part of the programme GRAVLENS25, we also observed \object{PS J0147+4630} with TTT3/COLORS from 12 September 2025 to 14 January 2026. On most observing nights, two 120 s exposures were taken in each of the $g$ and $r$ bands, while we performed an intra-night optical monitoring of the Andromeda Parachute in the period 10-14 January 2026 (see Sect.~\ref{sec:parachintra}). In the left panel of Figure~\ref{fig:parach_ima_spike}, a TTT3/COLORS $r$-band frame (FWHM = 0\farcs65) shows the compact arc formed by the brightest images A-C, and just below this, the faintest image D.  

Our GLENDAMA-GLAM observations of \object{PS J0147+4630} with LT/IO:O were performed from 22 July 2025 to 11 January 2026. The exposure time per frame was also 120 s, as with TTT3/COLORS, and the LT monitoring campaign 
overlapped with that from the TTT3 over a period of four months. Over the entire LT+TTT3 monitoring period spanning six months, we obtained usable data on 36 and 40 nights in the $g$ and $r$ bands, respectively. The TTT3 made a significant contribution to this database by providing information at 23 observing epochs in each band.

\subsection{Light curves}
\label{sec:parachphot}

In the strong lensing region (see the left panel of Figure~\ref{fig:parach_ima_spike}), the fluxes of the quasar images A-D were extracted through PSF fitting, using the IMFITFITS software \citep{McLe98}. These four images and the very faint early-type lens galaxy G were modelled as four empirical PSFs and a De Vaucouleurs distribution convolved with the empirical PSF, respectively \citep[details on the photometric method in][]{Goico19}. While this standard photometry worked well on the LT frames, the TTT3 frames contain a 12-mag saturated star (SS$_{\rm 12m}$) whose diffraction spikes sometimes contaminate the region of interest (see the right panel of Figure~\ref{fig:parach_ima_spike}). To remove this contamination in the sky region occupied by the lens system, we used the light distribution of another saturated field star as a template. The template was moved to the SS$_{\rm 12m}$ position, rescaled appropriately, and subtracted.   

In Figure~\ref{fig:parach_lcs}, and Tables 3 ($g$ band) and 4 ($r$ band) at the CDS, we present the LT+TTT3 light curves of \object{PS J0147+4630} and a non-variable field star S. To estimate the magnitude errors we followed a procedure similar to that described in the second paragraph of Sect.~\ref{sec:crossphot}, using adjacent epochs separated by no more than 4.5 d (instead of 2.5 d). Our photometric errors for the images A-D are very conservative because they could include a contribution due to the variability of the quasar on a timescale of a few days. Despite this possible variability contribution, we achieve a good photometric accuracy for the Andromeda Parachute (see the last two rows from the top in Figure~\ref{fig:quads_err}). In the $r$ band, the image flux uncertainty is typically $\sim$1\%, but it rises to $\sim$2.5\% for D using the LT. In the $g$ band, using the LT, the image flux uncertainty does not exceed 2.5\%. Additionally, the TTT3 $g$-band observations lead to a flux uncertainty of $\sim$1\% for A-C, but we only achieve 2.5$-$5\% photometry for the faintest quasar image. 

\subsection{Intra-night optical variability}
\label{sec:parachintra}

Non-lensed radio-quiet quasars can exhibit a certain level of intra-night optical variability \citep[e.g.,][]{Gupt05,Stal05,Negi23} that could be due to irradiation of the accretion disc by a variable X-ray flux, instability of the accretion disc, or the presence of a weak blazar component \citep[e.g.,][]{Czer08}. This intrinsic signal, if present in a GLQ, may be a critical tool to 
accurately determine the time delay(s) of the lens system \citep[e.g.,][]{Coll03}. 

We explored the intra-night variability of the Andromeda Parachute on some nights in January 2026. On each of these nights, the standard TTT3 observing sequence (2$\times$120 s $g$-band exposures plus 2$\times$120 s $r$-band exposures) was repeated multiple times during 1$-$3 hours. Although we obtained $\sim$20$-$30 usable frames in each of the two bands on January 10, 11, and 12, we did not find evidence of hourly or sub-hourly optical variability at amplitudes of $\geq$0.010 mag for images A-C. Unfortunately, our observations suggest that this quad does not exhibit the rapid events required to obtain highly precise measurements of its inter-image time delays. 

\section{\object{PS J0147+4630}: astrophysics from inter-image time delays}
\label{sec:parachvar}

\subsection{Time delays}

\begin{figure*}
\sidecaption
\includegraphics[width=6cm]{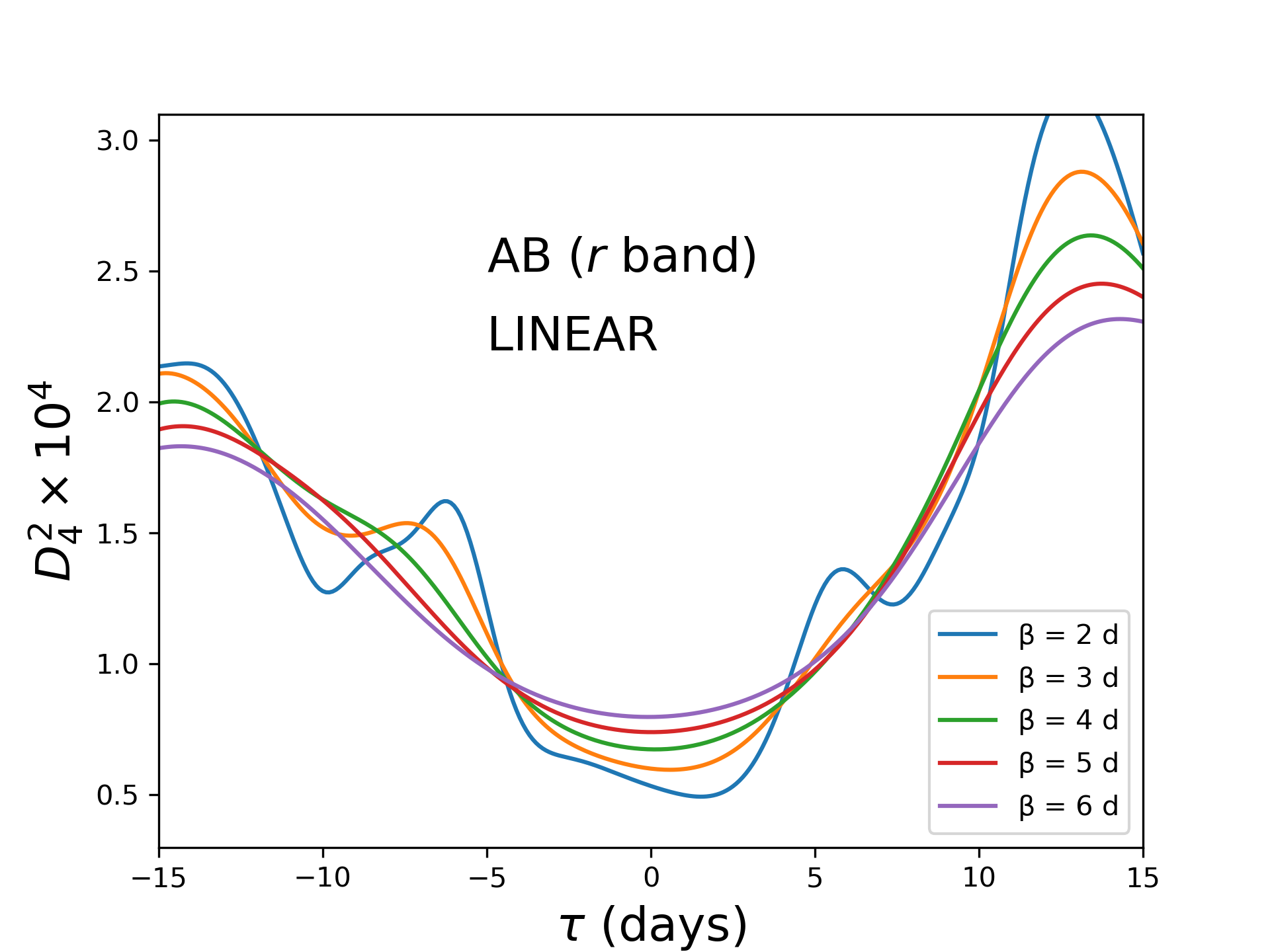} 
\includegraphics[width=6cm]{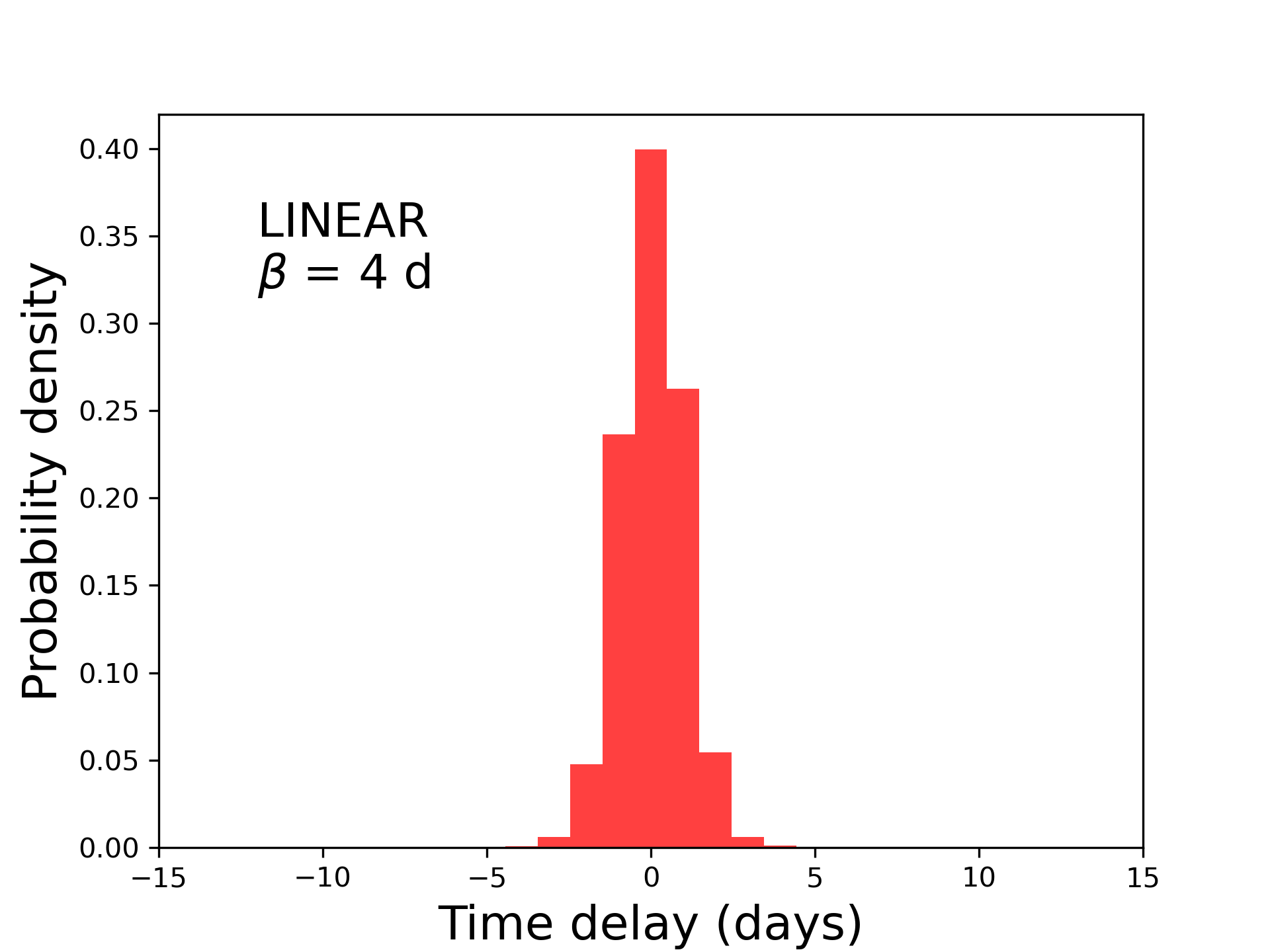}
\caption{AB delay estimation from LT+TTT3 $r$-band data and the dispersion technique. Left panel: dispersion spectra for linear microlensing and decorrelation lengths ($\beta$) ranging from 2 to 6 d. Right panel: time 
delay histogram from 10$^4$ pairs of simulated curves (A, B). The microlensing variability is assumed to follow a linear law and $\beta$ = 4 d.} 
\label{fig:parach_d2AB}
\end{figure*}

\setcounter{table}{4}
\begin{table*}[h!]
\begin{center}
\caption{Time delays between arc images in \object{PS J0147+4630}.}
\label{tab:ABCdelay}
\begin{tabular}{llcccc}
\hline\hline
Data & Method & \multicolumn{2}{c}{AB delay} & \multicolumn{2}{c}{AC delay}\\
\cline{3-4} \cline{5-6}
     &        & 1$\sigma$ & 2$\sigma$ & 1$\sigma$ & 2$\sigma$\\
\hline                                                       
NOT (2017$-$2019) &  $\chi^2$/Legendre polynomials [1]  & ... & $-$2.21$^{+2.08}_{-2.16}$ & ... & $-$5.28$^{+2.16}_{-2.22}$ [2]\\
LT (2017$-$2019) &  $D^2$/constant magnitude offset [3]  & $+$2.6$^{+3.2}_{-1.1}$ [4] & ... & ... & ...\\
LT+NOT (2017$-$2022) & $\chi^2$/free-knot splines [5-7] & $-$0.2 $\pm$ 3.9 & ... & $+$6.4 $\pm$ 4.4 [8] & ...\\
LT+TTT3 (2025$-$2026) [9]: & & & & & \\
$r$ band & $D^2$/linear microlensing   & 0.0 $\pm$ 0.9 & 0.0 $\pm$ 1.9 & $+$0.2 $\pm$ 0.9 & $+$0.2 $\pm$ 1.8\\
$g$ band & $D^2$/linear microlensing   & $-$0.8 $\pm$ 1.3 & $-$0.8 $\pm$ 3.1 & $-$0.4$^{+1.5}_{-1.6}$ & ...\\
\hline 
\end{tabular}
\end{center}
\footnotesize{Note: Here, we include all available measurements of $\Delta t_{\rm{AB}}$ and $\Delta t_{\rm{AC}}$ in days. The previous measurements were performed using data in the $rR$ red bands. The 1$\sigma$ and 2$\sigma$ confidence intervals in this paper (last two rows) are based on 10$^4$ simulated curves of A, B, and C, and a decorrelation length $\beta$ = 4 d (see the main text). The 2$\sigma$ confidence interval for the $g$-band AC delay spans $\sim$25 d and is not listed in the last row.}
\tablebib{
[1] \citet{Koch96}; 
[2] \citet{Dyrl19}; 
[3] \citet{Pelt96}; 
[4] \citet{Goico19};
[5] \citet{Tewe13}; 
[6] \citet{Mill20a};
[7] \citet{Mill20b};
[8] \citet{Shal23};
[9] This paper.
}
\end{table*} 

Recently, the inter-image time delays of \object{PS J0147+4630} were derived from $r$-band light curves based on observations with the LT and the 2.5 m Nordic Optical Telescope (NOT) in the 2017$-$2022 period \citep{Shal23}. While the delays between image D and the other three images were measured to be in the range from 170 to 177 d with uncertainty of $\sim$4\%, only weak constraints were obtained on the short delays between the brightest images: $\Delta t_{\rm{AB}}$ = $-$0.2 $\pm$ 3.9 d and $\Delta t_{\rm{AC}}$ = $+$6.4 $\pm$ 4.4 d (1$\sigma$ confidence intervals), where $\Delta t_{\rm{AX}} = t_{\rm{A}} - t_{\rm{X}}$. To try to improve these constraints on the AB and AC delays, in the absence of intra-night variations (see Sect.~\ref{sec:parachintra}), we mainly focused on the LT+TTT3 $r$-band data in the right panel of Figure~\ref{fig:parach_lcs}. The analogy between the ABC curves indicates the presence of intrinsic quasar variability and weak microlensing gradients. 

To cross-correlate the light curves of A and B (C), we used the $D^2_{4,2}$ dispersion technique \citep{Pelt96}. This well-tested method allows us to find the AB (AC) delay and a possible microlensing signal, providing the 
information required to match the two records. We assumed that the slow microlensing signal can be modelled as a linear polynomial. We also considered a Gaussian weighting scheme characterised by a decorrelation length $\beta$ \citep[e.g., see Appendix A of][]{Shal25}. 

In the left panel of Figure~\ref{fig:parach_d2AB}, we show dispersion spectra for the AB pair and five values of $\beta$. For example, $\beta$ = 4 d (green line) produces a smooth spectrum having a minimum at a time lag $\tau$ = $+$0.1 d, which is a rough estimation of $\Delta t_{\rm{AB}}$. There is also a very small microlensing gradient of $-$3.2 $\times$ 10$^{-5}$ mag d$^{-1}$, resulting in a microlensing-induced variation of 6 mmag in 180 days. We adopted $\beta$ = 4 d as a compromise between smoothness of the dispersion spectrum and depth of its minimum, avoiding both noisy (low $\beta$) and over-smoothed (high $\beta$) behaviours. To obtain the corresponding confidence intervals, we generated 10$^4$ simulated light curves of each quasar image at epochs equal to those of observation, and then obtained the $D^2_{4,2}$ spectrum and its minimum for all (A, B) pairs of the simulated curves \citep[e.g.,][]{Shal25}. The distribution of the time delays for the 10$^4$ repetitions of the experiment is depicted in the right panel of Figure~\ref{fig:parach_d2AB}, and Appendix~\ref{sec:histdelay} shows how the $r$-band light curves of A and B match using different delays.

Following the same scheme for the AC image pair, i.e., using the $r$-band data, linear microlensing, $\beta$ = 4 d, and 10$^4$ simulated light curves of A and C, we derived a time delay histogram similar to that in the right panel of Figure~\ref{fig:parach_d2AB}. Our AB and AC delay measurements in both optical bands are presented and compared with previous ones in Table~\ref{tab:ABCdelay}, and from now on we only consider the $r$-band delays because the $g$-band data provide weaker constraints. The new 2$\sigma$ measurement of the AB delay is consistent with the constraint of \citet{Dyrl19}, but it does not rule out positive values (image B leading image A). In addition, the 2$\sigma$ confidence intervals for the AC delay do not agree with each other. Although this difference might be due to variability at different points of a finite-size source \citep{Yone99} or to a microlensing-induced change in the AC delay \citep{Tie18}, in order to shed light on the lensing mass distribution, we considered the new measurement as it is. 

\subsection{Mass modelling}

\begin{table*}[h!]
\begin{center}
\caption{Inter-image delays predicted by the SPLE + ES lens model.}
\label{tab:predelay}
\begin{tabular}{llccc}
\hline\hline
Software & Constraints & AB delay & AC delay & AD delay\\
\hline                                                       
LENSTRONOMY &  HST imaging & $-$2.7 $\pm$ 0.4 & $-$8.8 $\pm$ 1.3 & $-$244 $\pm$ 23\\
GRAVLENS/LENSMODEL & HST astrometry \& LT+TTT3 delays & $+$1.4 & $+$1.4 & $-$187.2\\
LENSTRONOMY & HST astrometry \& parameter priors & $+$1.58$^{+0.15}_{-0.14}$ & $+$1.17$^{+0.12}_{-0.10}$ & $-$158.2$^{+5.4}_{-6.3}$\\
\hline 
\end{tabular}
\end{center}
\footnotesize{Note: All best values and 1$\sigma$ confidence intervals are in days. Delays in the second row correspond to those in Table C1 of \citet{Shaj19} after correcting for updated redshifts. Delays in the last two rows are results in this paper (see the main text).}
\end{table*} 

The lensing mass distribution is usually modelled through a singular power-law ellipsoid (SPLE) that describes the main lensing galaxy G, and an external shear (ES) that accounts for the gravitational effect of additional galaxies along the line of sight to the system \citep[e.g.,][]{Suyu13,Shaj19}. If these additional galaxies produce an external convergence $\kappa_{\rm ext}$, some SPLE + ES model parameters must be reinterpreted \citep[e.g.,][and references therein]{Shal25}. 

\citet{Shaj19} used the LENSTRONOMY software\footnote{https://github.com/sibirrer/lenstronomy} \citep{Birr15,Birr18}, the SPLE + ES model and Hubble Space Telescope (HST) imaging to reconstruct the lensing mass of \object{PS J0147+4630} through an automated modeling scheme. From updated redshifts of the lens system \citep[$z_{\rm s}$ = 2.357 and $z_{\rm l}$ = 0.678;][]{Goico19,Goico26} and a concordance flat cosmology with Hubble constant $H_0$ = 70 km s$^{-1}$ Mpc$^{-1}$, Shajib et al.'s mass solution yields time delays between arc images of $-$2.7 d (AB) and $-$8.8 d (AC), which are close to the central values of the NOT delay intervals in Table~\ref{tab:ABCdelay}, but differ appreciably from those of the LT+TTT3 intervals. Assuming that the SPLE + ES scenario provides a good starting point for describing the actual mass distribution, the average delay ratio $\Delta t^{\rm{NOT}} / \Delta t^{\rm{model}} \sim$ 0.7 suggests $\kappa_{\rm ext} \sim$ 0.3 \citep[e.g., see Equation (16) of][]{Birr24}. This very high external convergence also arises when calculating the ratio between the central value of the NOT AD delay interval \citep[$-$170.5 d;][]{Dyrl19} and the AD delay from the mass solution ($-$244 d; see Table~\ref{tab:predelay}). 

To investigate the physics behind the AB and AC delays from our LT+TTT3 short-term monitoring, we also used the GRAVLENS/LENSMODEL software\footnote{http://www.physics.rutgers.edu/~keeton/gravlens/} \citep{Keet01,Keet04} and the SPLE + ES model, treating the HST astrometry \citep{Shaj19} and the 2$\sigma$ delay intervals in the second to last row of Table~\ref{tab:ABCdelay} as observational constraints. We found a best solution with power-law slope $\gamma$ = 1.9, which is lower than the isothermal value ($\gamma$ = 2.0) derived from \citet{Shaj19}. Furthermore, although Shajib et al.’s mass reconstruction indicated that the centre of the SPLE is displaced from the the light centroid of G by approximately 0\farcs13 to the South, our best solution corresponds to a positional offset of only $\sim$0\farcs07 to the West. Additionally, the predicted AB and AC delays are positive, and the AD delay in the third row of Table~\ref{tab:predelay} is close to the measured value (see above).

As an additional test, we also carried out blindly an analysis of \object{PS J0147+4630} using LENSTRONOMY in its modelling catalogue mode. We only took the astrometry from the HST and the SPLE + ES model\footnote{SPLE is called PEMD in LENSTRONOMY nomenclature} to obtain a new mass solution and its associated time delays, assuming updated redshifts and a fiducial cosmology as in the GRAVLENS/LENSMODEL case. Since previous modelling results \citep{Berg17,Shaj19} provided well-concordance values in $\theta_{\rm E}$ (Einstein angular radius of the SPLE) and $\gamma_{\rm ext}$ (ES strength), we adopted Gaussian priors on those parameters for the model: $\theta_{\rm E}$ = 1\farcs9 with $\sigma(\theta_{\rm E})$ = 0\farcs1 and $\gamma_{\rm ext, 1}$ = 0.12 with $\sigma(\gamma_{\rm ext, 1})$ = 0.1. In addition, the centre of the SPLE was allowed to be displaced from the the light centroid of G by $\leq$0\farcs01. These priors were crucial to successfully guide the model to a SPLE slope different from a pure isothermal value when evaluating the likelihood. 

The results for the median values and confidence intervals of the model parameters are plotted as covariance matrices in Figure \ref{fig:cornerplot}, and the 1$\sigma$ confidence intervals for the time delays were calculated from the last 20\% MCMC samples. We note that the mass distribution of the SPLE is shallower than isothermal ($\gamma=1.82$), in good agreement with the best solution from the GRAVLENS/LENSMODEL software based on the HST astrometry and the LT+TTT3 delays between arc images. Notably, the predicted time delays in the fourth row of Table~\ref{tab:predelay} are similar in sign and amount to those from the GRAVLENS/LENSMODEL solution. Additionally, using the two new mass solutions, the average predicted AD delay of $-$172.7 d indicates a usual value of the external convergence of 0.013 \citep[e.g.,][]{TDCOSMO25}.

\section{Conclusions}
\label{sec:end}

Short-term optical monitoring campaigns of GLQs may be used to explore their astrophysical potential and thus adequately plan long-term follow-ups. Regarding short-term monitoring, galaxy-scale quads are particularly relevant because they have some inter-image time delay (if not all) of the order of hours, days or weeks \citep[e.g.,][]{Duco26}. As proof of concept, over a period of about six months, we monitored the optical variability of two known quads using the 2m-class robotic telescopes LT and TTT3. The observational programme consisted of exposures in the $g$ and $r$ bands. 

For the Einstein Cross (\object{QSO 2237+0305}), X-ray variability studies led to inter-image time delays of the order of hours \citep[e.g.,][]{Dai03,Fedo26}, which agree with lens model predictions \citep[e.g.,][]{Schm98}. Therefore, using optical light curves with a standard cadence of one day or longer, differences between two quasar images reveal microlensing-induced variations, and thus information about the physics of the lensed source and the lensing galaxy \citep[e.g.,][]{Vern24}. From our LT+TTT3 $g$-band data, we constructed the difference light curve between images A and B. Adopting reasonable physical properties of the lensing galaxy, this difference signal and numerical microlensing simulations were then used to constrain the half-light radius $R_{\rm s}$ of the $g$-band source. We obtained a range of $R_{\rm s}$ values in excellent agreement with results from 12 years of monitoring in the $g$ band \citep{Shal26}, validating the use of the LT+TTT3 data to probe the physics of \object{QSO 2237+0305} and to prepare long-term observational plans \citep[e.g., a muti-band monitoring to discuss the source radius-wavelength relationship, which has already been successfully implemented;][]{Shal26}.   

For the Andromeda Parachute (\object{PS J0147+4630}), TTT3 few-minute cadence light curves showed no evidence of intra-night variability, so fast optical events cannot be used to accurately determine the delays between ABC images in the bright arc of the lens system. However, the LT+TTT3 $r$-band light curves with a standard cadence allowed us to obtain valuable estimates of the AB and AC time delays with 2$\sigma$ uncertainties of $\sim$2 d. We should note that the new 2$\sigma$ measure of the AC delay differs significantly from the 2$\sigma$ interval from a previous short-term monitoring campaign \citep{Dyrl19}. In addition, using HST imaging to constrain the SPLE + ES lens model, the mass solution is consistent with previous inter-image delays and a very high external convergence. However, using the HST astrometry and the new delays between arc images as constraints (GRAVLENS/LENSMODEL case) or astrometric postions with some priors in the parameters as a first approximation (LENSTRONOMY with catalogue data case), the resulting solutions favour a SPLE a little shallower than isothermal, leading to similar predictions for the inter-image delays and to an external convergence of order 10$^{-2}$. In the future, dense monitoring should be used to accurately determine whether A is trailing or leading images BC. 

Optical monitoring of quads with ground-based 2m-class telescopes for 150$-$180 days enables the assessment of their astrophysical relevance. Data can be distributed unevenly over the monitoring period, with a mean observing cadence of 4$-$5 days and denser sampling in some time segments. A few-minute cadence on selected nights also allow us to check for the presence of intra-night variability. Atmospheric turbulence causes blurring in ground-based astronomical imaging (FWHM $\sim$ 1\arcsec), limiting the spatial resolution of lens systems. Thus, quads with images separated by $\ga$1\arcsec can be observed with standard telescopes on the ground\footnote{If the lensing galaxy has a bright nucleus, this must also be separated by $\ga$1\arcsec from the quasar images.}, while the most compact systems require the use of additional instruments for improving spatial resolution or space telescopes such as the HST. The image brightness also influences the choice of a suitable telescope size, so ground-based 2m-class telescopes are well suited for monitoring systems with components brighter than 20 mag. Using these facilities, $\sim$2-minute exposures lead to $\sim$1\% photometry for images with $r$ = 16$-$16.5 mag in crowded regions. In addition, $\sim$3-minute exposures bring $\sim$2.5$-$5\% photometry for close-together components with $r$ = 18$-$18.5 mag each.

\section*{Data availability}

Tables 1$-$4 are only available in electronic form at the CDS via anonymous ftp to cdsarc.u-strasbg.fr (130.79.128.5) or via http://cdsweb.u-strasbg.fr/cgi-bin/qcat?J/A+A/vol/id
     
\begin{acknowledgements}
We thank the referee for several valuable comments that helped us improve the original manuscript. This article is based on observations made in the Two-meter Twin Telescope (TTT) sited at the Teide Observatory of the Instituto de Astrofísica de Canarias (IAC), that Light Bridges operates in Tenerife, Canary Islands (Spain). The observation time rights used for this research were consumed in the PEI 'GRAVLENS25'. This article used flash storage and GPU computing resources as Indefeasible Computer Rights (ICRs) being commissioned at the ASTRO POC project that Light Bridges will operate in Tenerife. The ICRs were consumed in the PEI 'GRAVLENS25' with the collaboration of Bechtle and LENOVO. The Liverpool Telescope (LT) is operated on the island of La Palma by Liverpool John Moores University in the Spanish Observatorio del Roque de los Muchachos of the IAC with financial support from the UK Science and Technology Facilities Council. We thank the staff of the LT and TTT for a kind interaction. This research was supported by funds of the Universidad de Cantabria and the grant PID2020-118990GB-I00 funded by MCIN/AEI/10.13039/501100011033. 
\end{acknowledgements}

\begin{appendix}

\onecolumn
\vspace{-1.5cm}
\section{Photometric accuracy from the LT and TTT3 observations}
\label{sec:photaccu}
\centering
\includegraphics[width=0.3\textwidth, trim=5 0 35 15, clip]{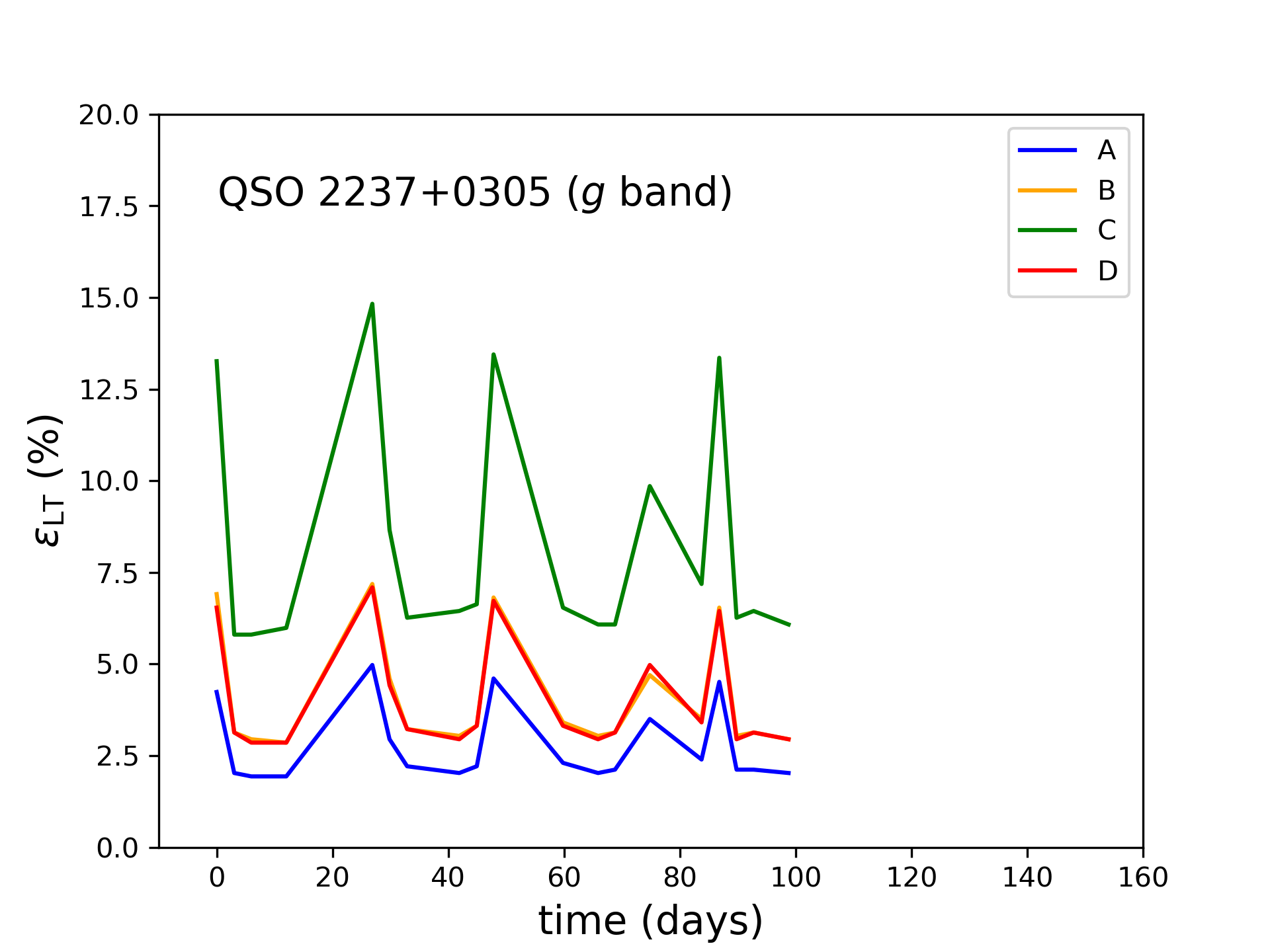}
\includegraphics[width=0.3\textwidth, trim=5 0 35 15, clip]{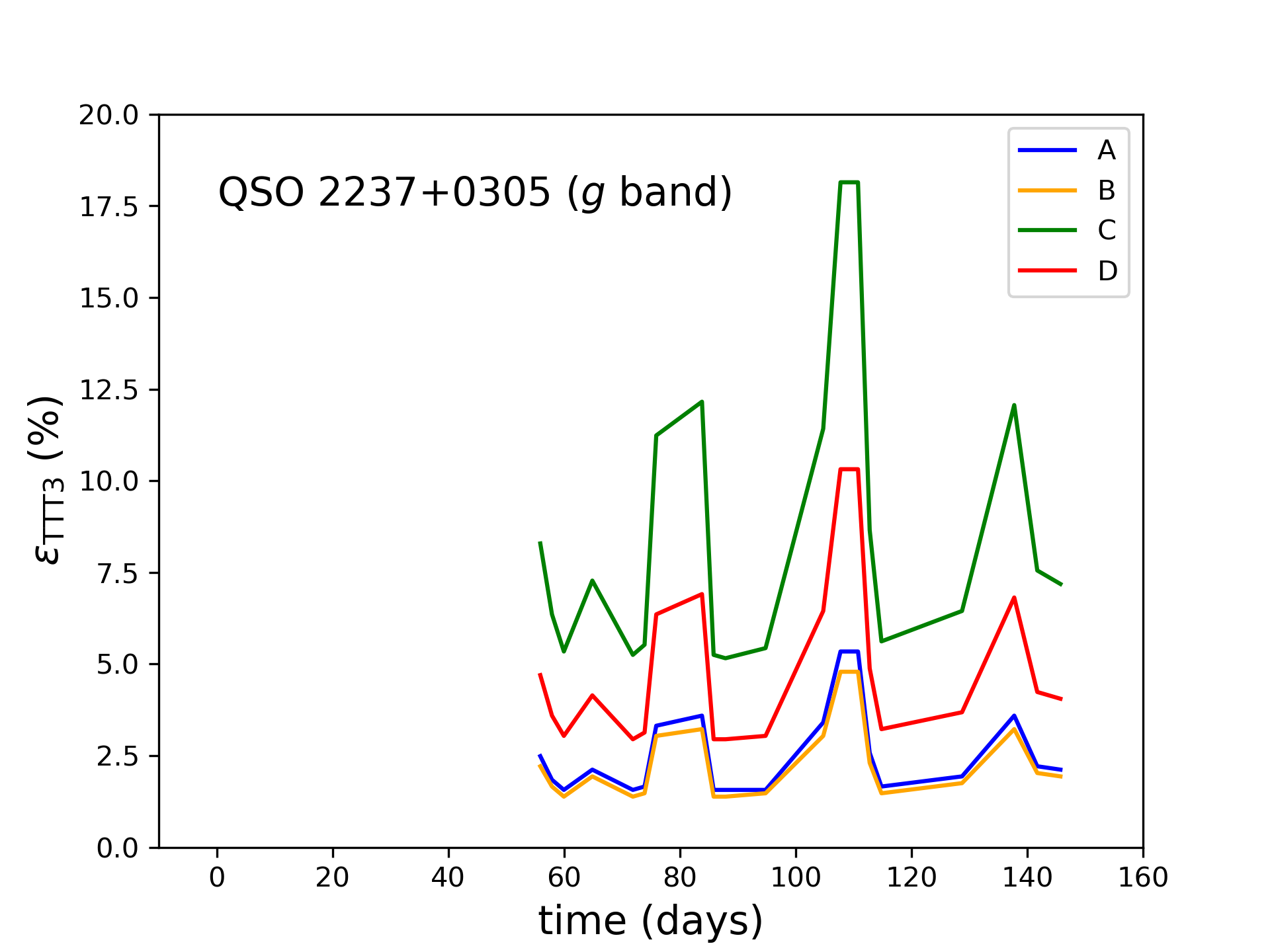}
\includegraphics[width=0.3\textwidth, trim=5 0 35 15, clip]{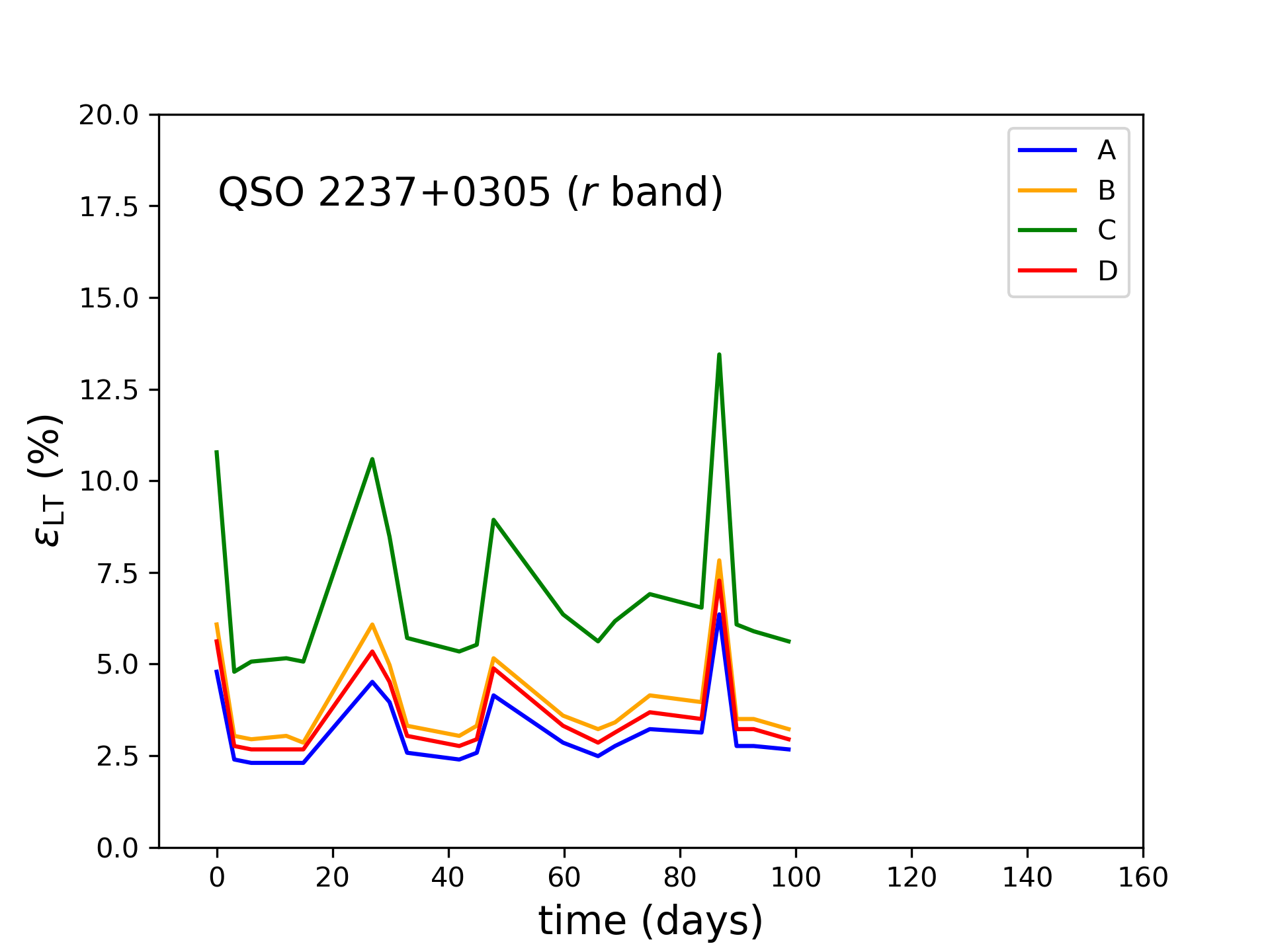}
\includegraphics[width=0.3\textwidth, trim=5 0 35 15, clip]{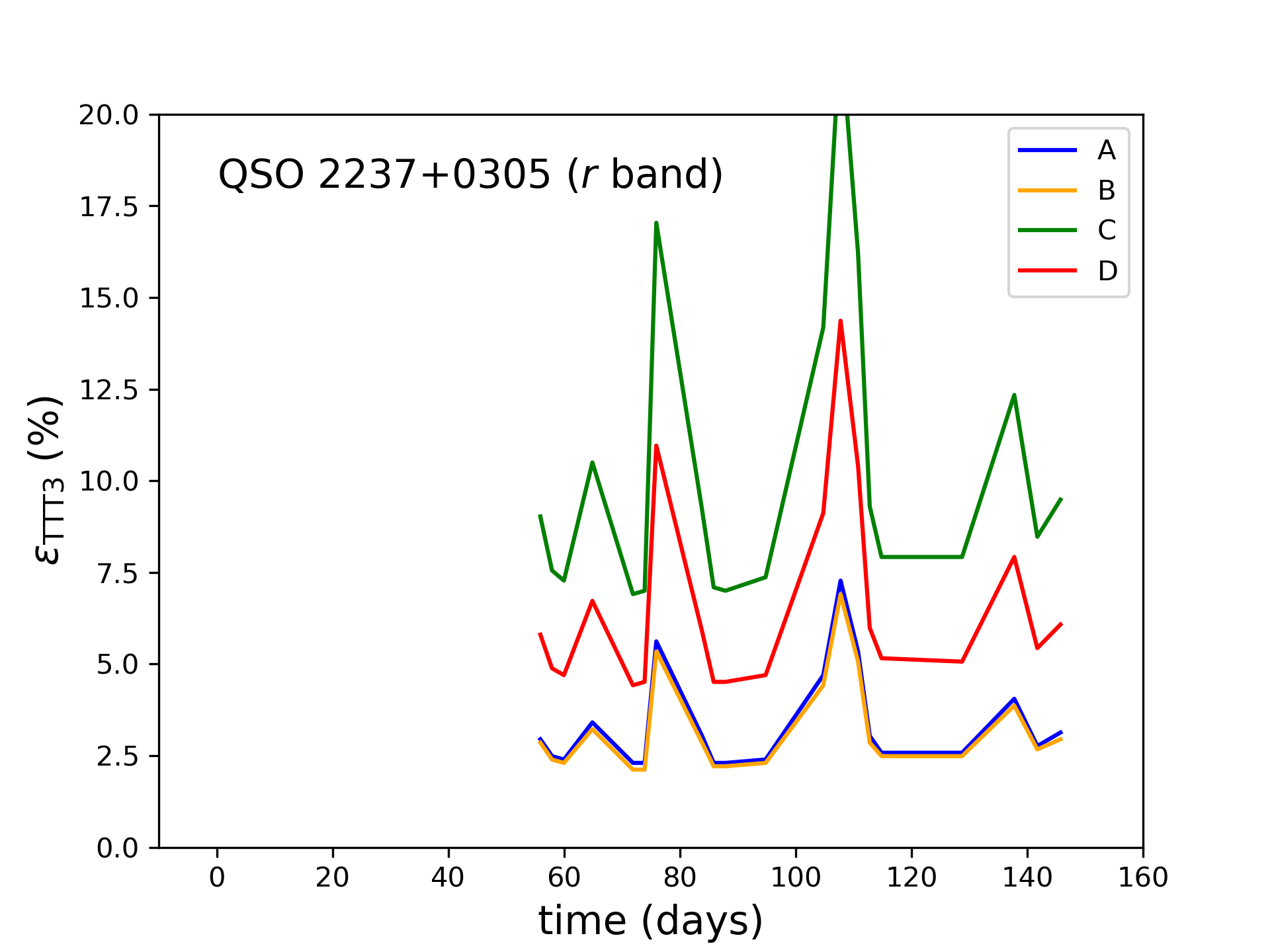}
\includegraphics[width=0.3\textwidth, trim=5 0 35 15, clip]{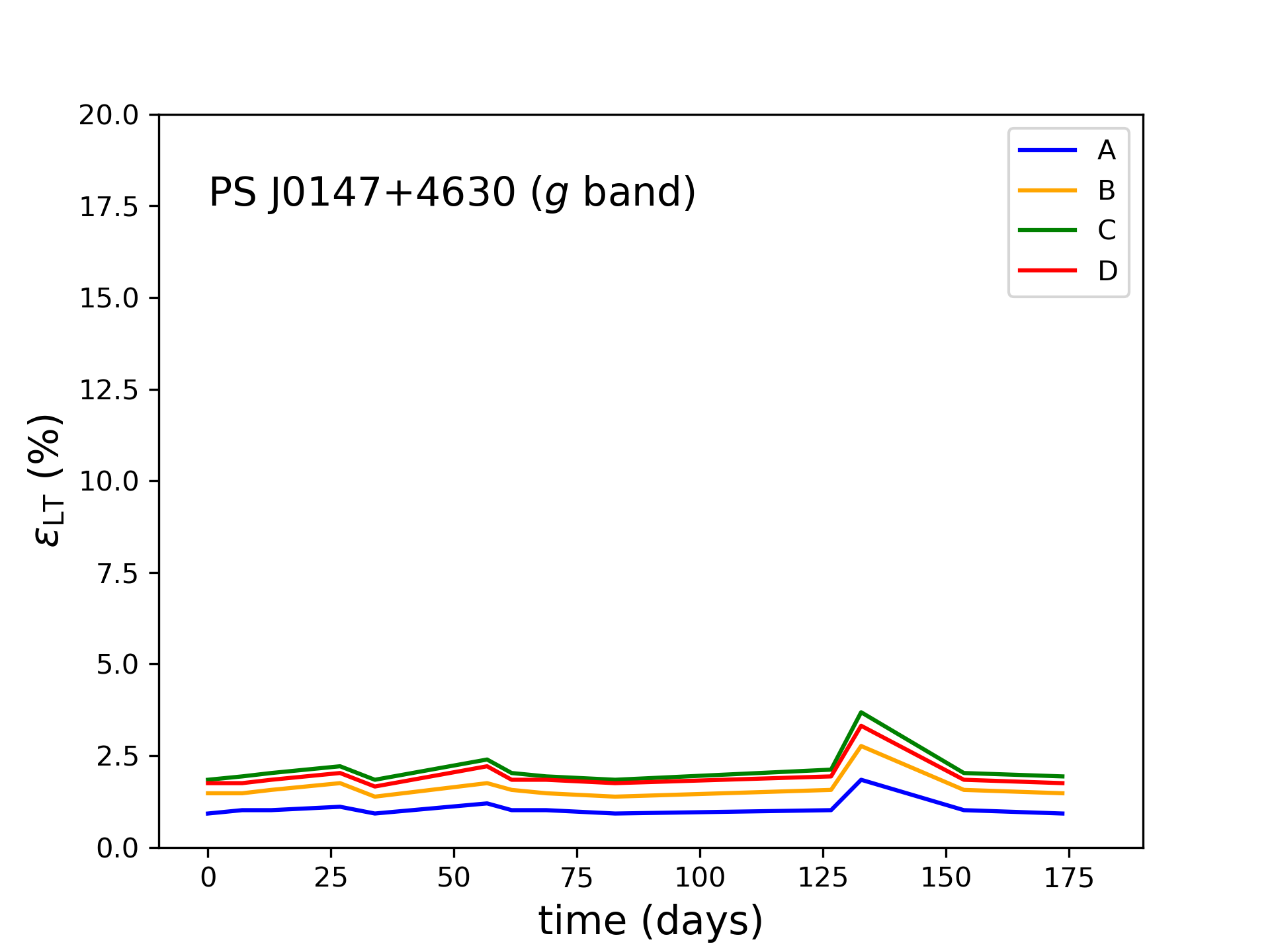}
\includegraphics[width=0.3\textwidth, trim=5 0 35 15, clip]{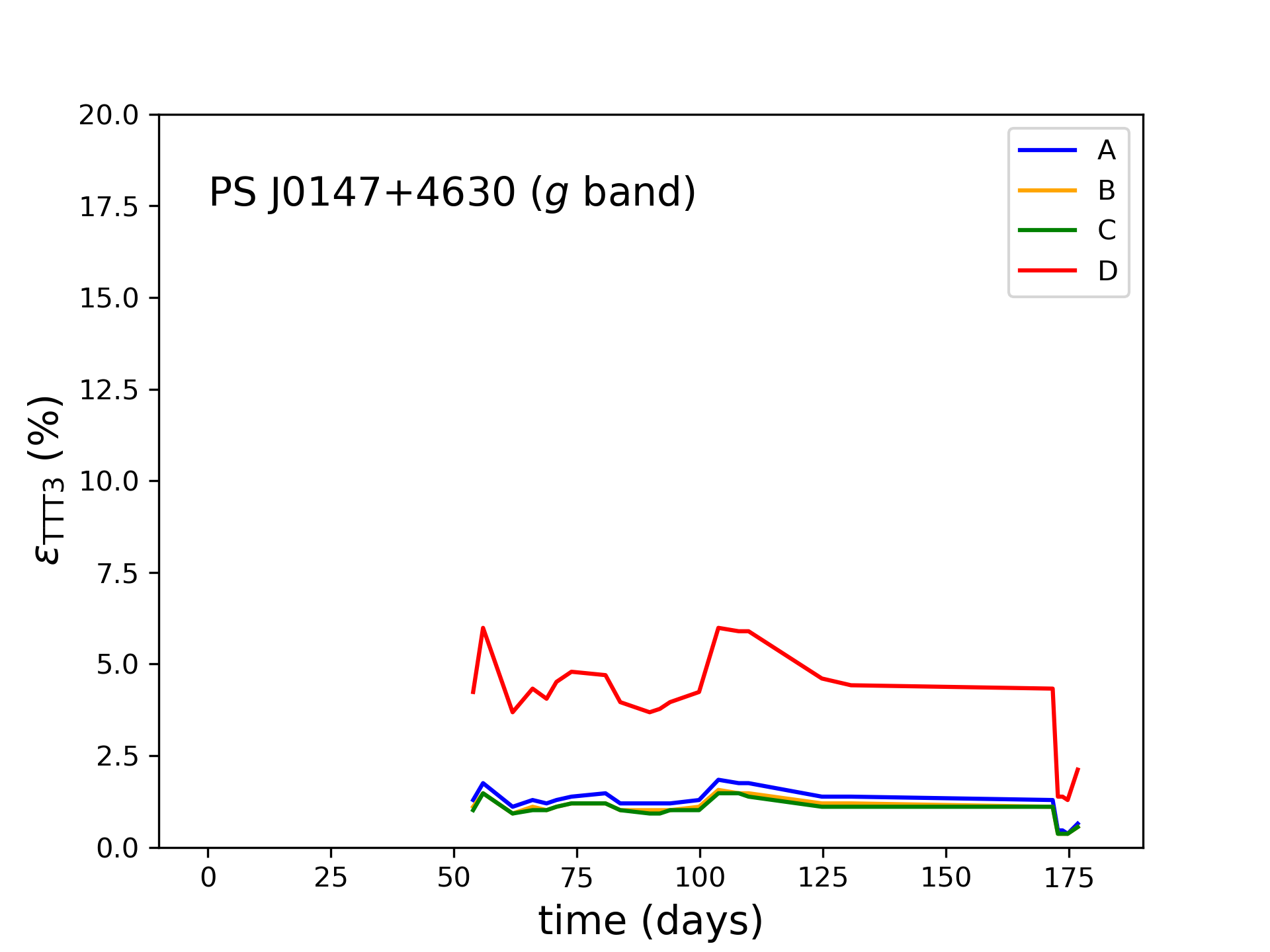}
\includegraphics[width=0.3\textwidth, trim=5 0 35 15, clip]{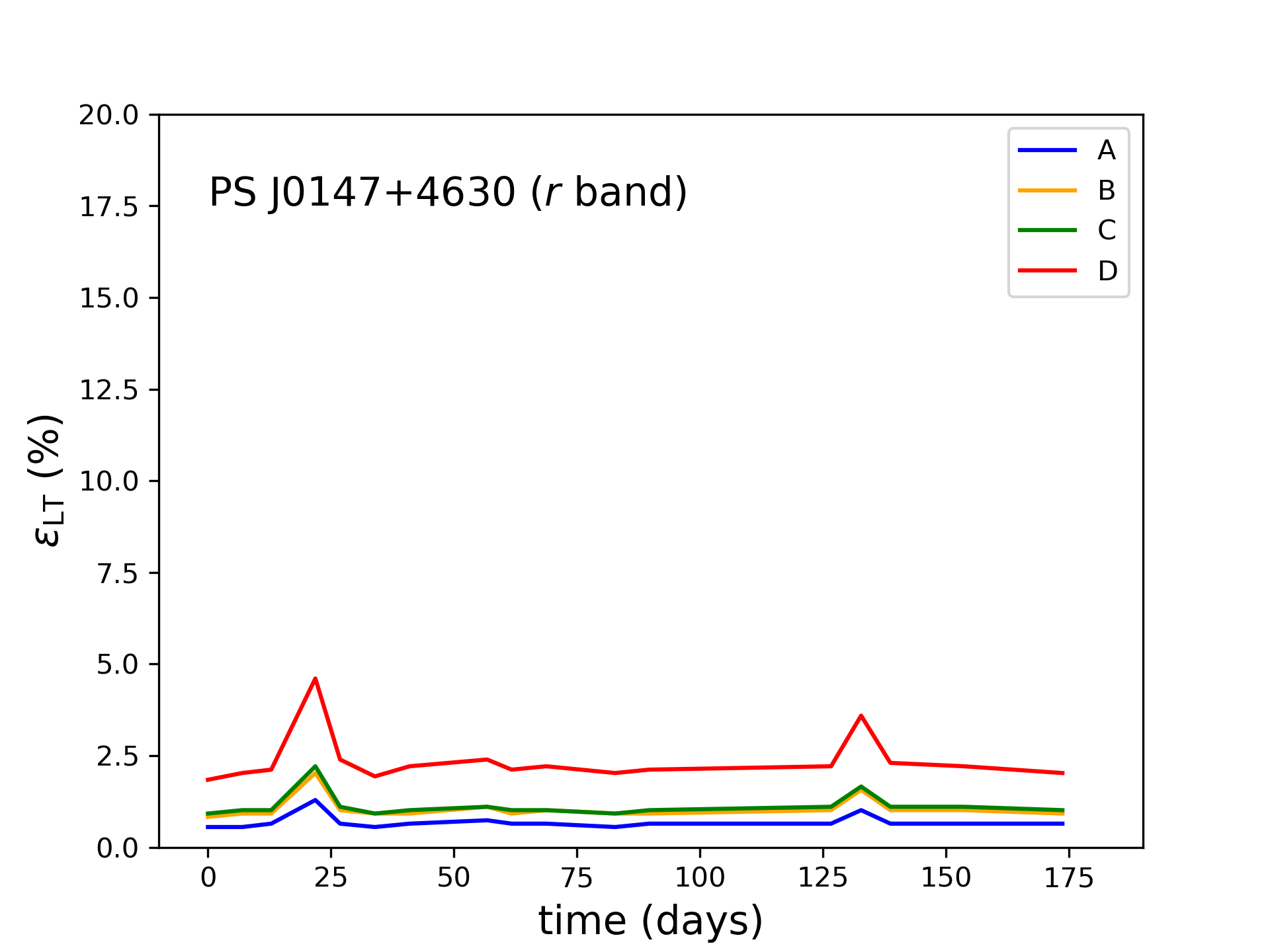}
\includegraphics[width=0.3\textwidth, trim=5 0 35 15, clip]{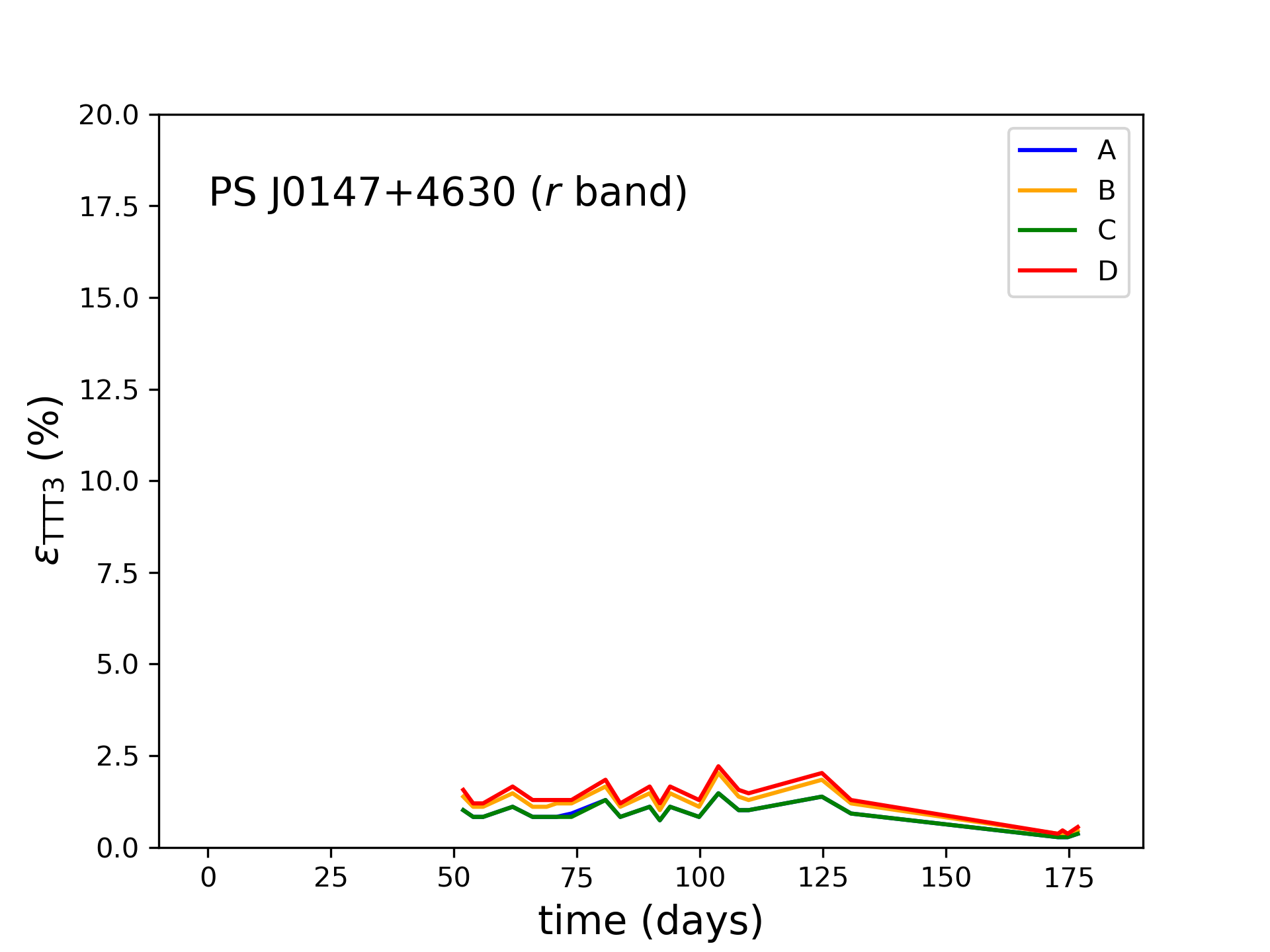}
\captionof{figure}{Percent flux uncertainty. The $y$-axis denotes the results from LT and TTT3 observations, for each system and band.}
\label{fig:quads_err}

\FloatBarrier
\clearpage
\section{Inter-telescope calibration}
\label{sec:intertelcal} 
 \justifying For each quad and optical band, constant magnitude offsets between the LT and TTT3 light curves were calculated using N$_{\rm p}$ LT-TTT3 pairs of magnitudes with time separations $\leq \delta$. For each quasar image, we estimated the mean value of the magnitude differences LT $-$ TTT3 (MEAN) and its standard error (SEM). We also considered the median of the differences as a reliable magnitude offset (MOFF) that minimises the impact of possible outliers. In the absence of outliers, MOFF $\sim$ MEAN and the SEM is a good proxy for the uncertainty in MOFF. Our results are shown in Tables~\ref{tab:crossg_twotel}, ~\ref{tab:crossr_twotel}, ~\ref{tab:parachg_twotel}, and ~\ref{tab:parachr_twotel}. The combined quasar records in Figures~\ref{fig:cross_lcs} and \ref{fig:parach_lcs} rely on the LT light curves and the TTT3 curves vertically shifted by the MOFF values. In all cases where no outliers are involved, the SEM values are significantly smaller than the typical magnitude errors of the TTT3 data, so magnitude offset uncertainties can be neglected. 

\begin{table}[!h]
\begin{center}
\caption{Inter-telescope calibration of \object{QSO 2237+0305} in the $g$ band (N$_{\rm p}$ = 7 and $\delta$ = 1.5 d).}
\label{tab:crossg_twotel}
\begin{tabular}{lccc}
\hline\hline
Image & MEAN & SEM & MOFF \\
\hline   
A & $+$0.072 & 0.007 & $+$0.067\\ 
B & $-$0.010 & 0.023 & $+$0.025$^{\ast}$\\ 
C & $-$0.075 & 0.017 & $-$0.075\\ 
D & $+$0.083 & 0.013 & $+$0.099\\
\hline 
\end{tabular}
\end{center}
\footnotesize{Note: An asterisk in the MOFF value indicates the presence of outliers in the magnitude differences LT $-$ TTT3.} 
\end{table} 
\vspace{-0.9em}
\begin{table}[!h]
\begin{center}
\caption{Inter-telescope calibration of \object{QSO 2237+0305} in the $r$ band (N$_{\rm p}$ = 7 and $\delta$ = 1.5 d).}
\label{tab:crossr_twotel}
\begin{tabular}{lccc}
\hline\hline
Image & MEAN & SEM & MOFF \\
\hline   
A & $+$0.041 & 0.010 & $+$0.041\\ 
B & $+$0.003 & 0.010 & $-$0.008\\ 
C & $+$0.007 & 0.043 & $+$0.018$^{\ast}$\\ 
D & $+$0.034 & 0.026 & $+$0.025\\
\hline 
\end{tabular}
\end{center}
\footnotesize{Note: An asterisk in the MOFF value indicates the presence of outliers in the magnitude differences LT $-$ TTT3.} 
\end{table} 
\vspace{-0.9em}
\begin{table}[!h]
\begin{center}
\caption{Inter-telescope calibration of \object{PS J0147+4630} in the $g$ band (N$_{\rm p}$ = 19 and $\delta$ = 5.5 d).}
\label{tab:parachg_twotel}
\begin{tabular}{lccc}
\hline\hline
Image & MEAN & SEM & MOFF \\
\hline   
A & $+$0.101 & 0.003 & $+$0.103\\ 
B & $+$0.105 & 0.003 & $+$0.105\\ 
C & $+$0.087 & 0.004 & $+$0.087\\ 
D & $+$0.028 & 0.008 & $+$0.020\\
\hline 
\end{tabular}
\end{center} 
\end{table} 
\vspace{-0.9em}
\begin{table}[!h]
\begin{center}
\caption{Inter-telescope calibration of \object{PS J0147+4630} in the $r$ band (N$_{\rm p}$ = 22 and $\delta$ = 5.5 d).}
\label{tab:parachr_twotel}
\begin{tabular}{lccc}
\hline\hline
Image & MEAN & SEM & MOFF \\
\hline   
A & $-$0.014 & 0.002 & $-$0.015\\ 
B & $-$0.012 & 0.003 & $-$0.015\\ 
C & $-$0.026 & 0.004 & $-$0.027\\ 
D & $-$0.043 & 0.004 & $-$0.048\\
\hline 
\end{tabular}
\end{center} 
\end{table} 
\FloatBarrier
\clearpage
\section{Examples of magnification maps for images A and B of \object{QSO 2237+0305}}
\label{sec:magmaps} 
\centering

\includegraphics[width=0.25\textwidth, trim=5 5 35 5, clip]{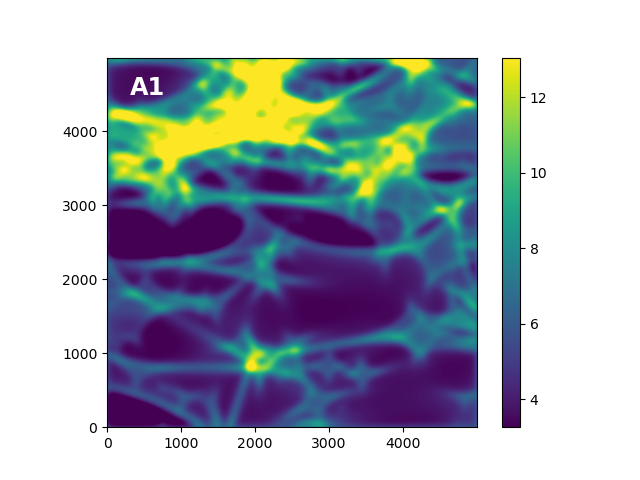}%
\includegraphics[width=0.25\textwidth, trim=5 5 35 5, clip]{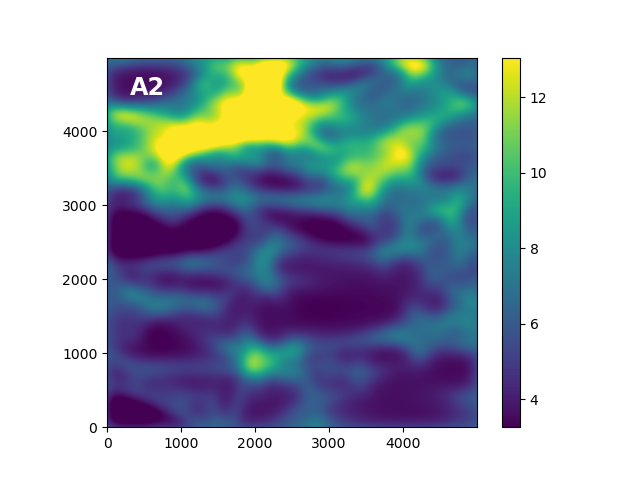}%
\includegraphics[width=0.25\textwidth, trim=5 5 35 5, clip]{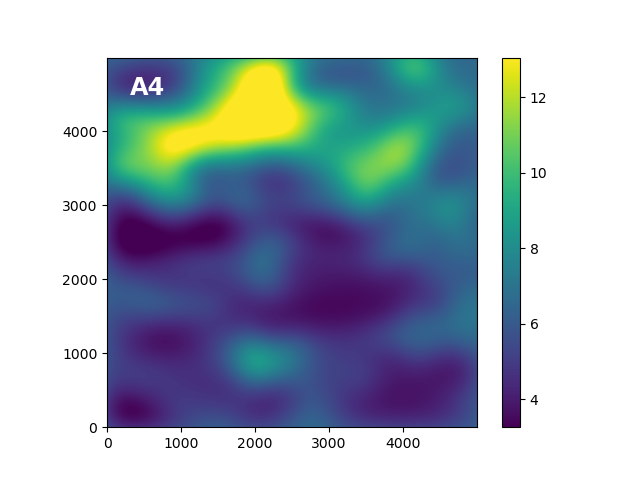}%
\includegraphics[width=0.25\textwidth, trim=5 5 35 5, clip]{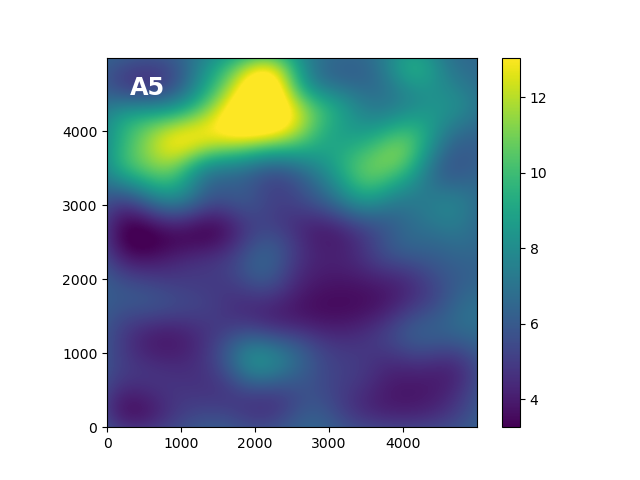}
\vspace{-0.15cm}

\includegraphics[width=0.25\textwidth, trim=5 5 35 5, clip]{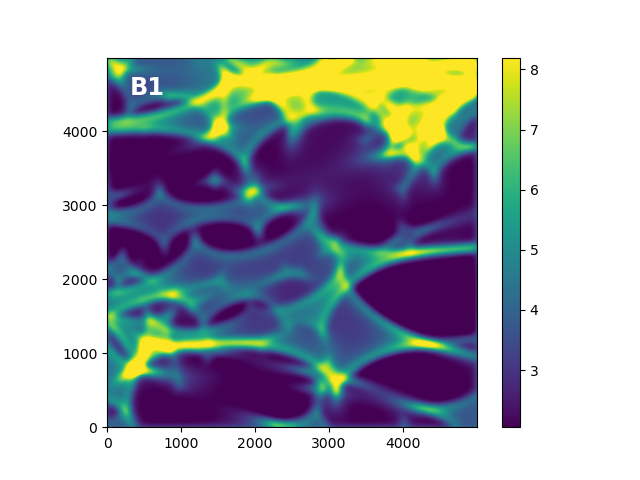}%
\includegraphics[width=0.25\textwidth, trim=5 5 35 5, clip]{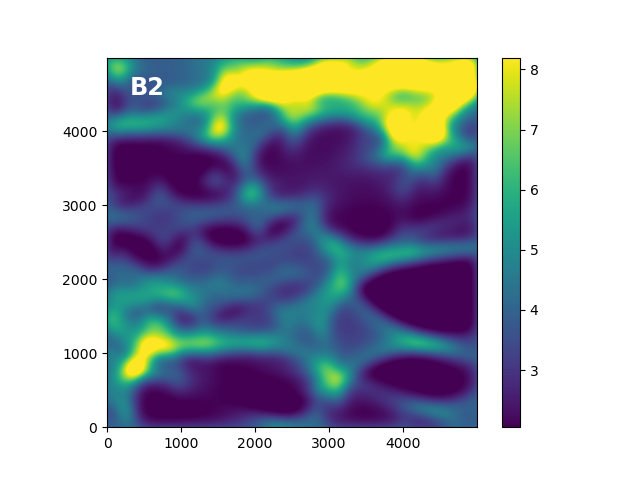}%
\includegraphics[width=0.25\textwidth, trim=5 5 35 5, clip]{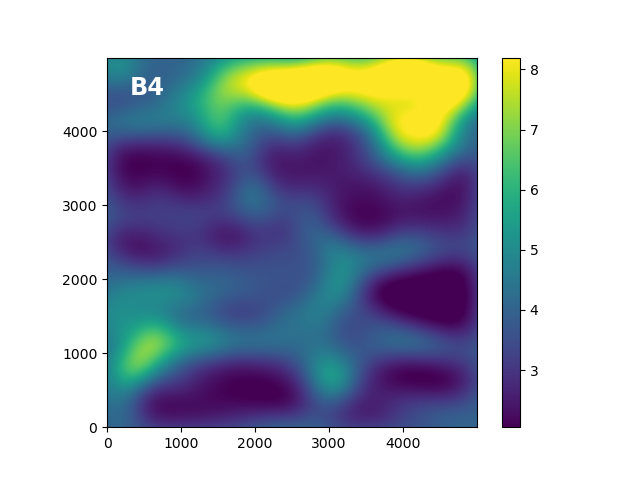}%
\includegraphics[width=0.25\textwidth, trim=5 5 35 5, clip]{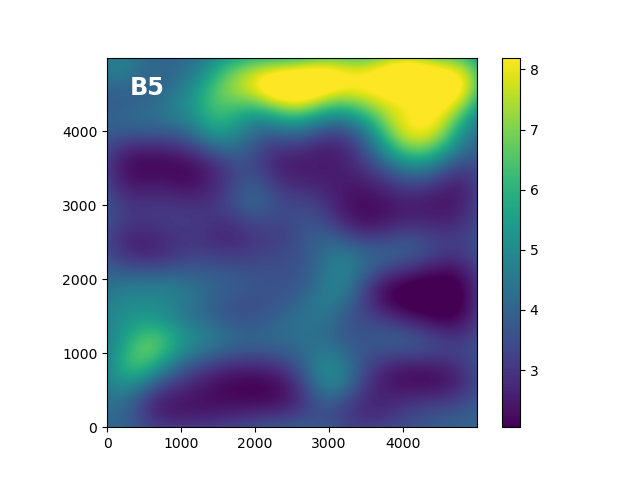}
\captionof{figure}{Magnification maps for the second analysis. The numerical label of each map (between 1 and 5) represents ten times the scaled radius $R/R_{\rm E}$ of the Gaussian source.}
\label{fig:cross_mmaps2}

\section{Time delay between images A and B of \object{PS J0147+4630}}
\label{sec:histdelay} 
\begin{figure*}[h!]
\centering
\includegraphics[width=0.3\textwidth, trim=2 0 35 5, clip]{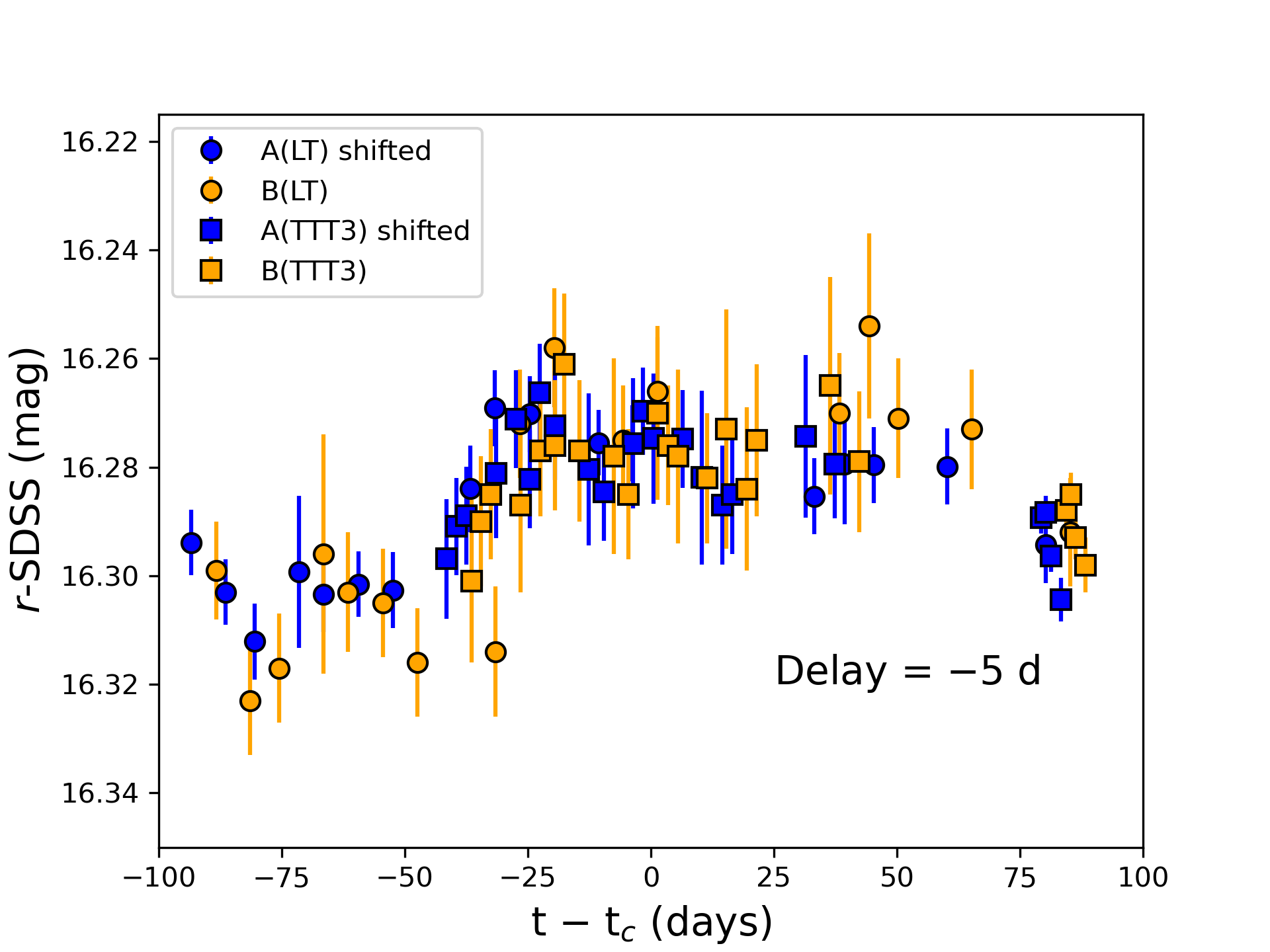}
\includegraphics[width=0.3\textwidth, trim=2 0 35 5, clip]
{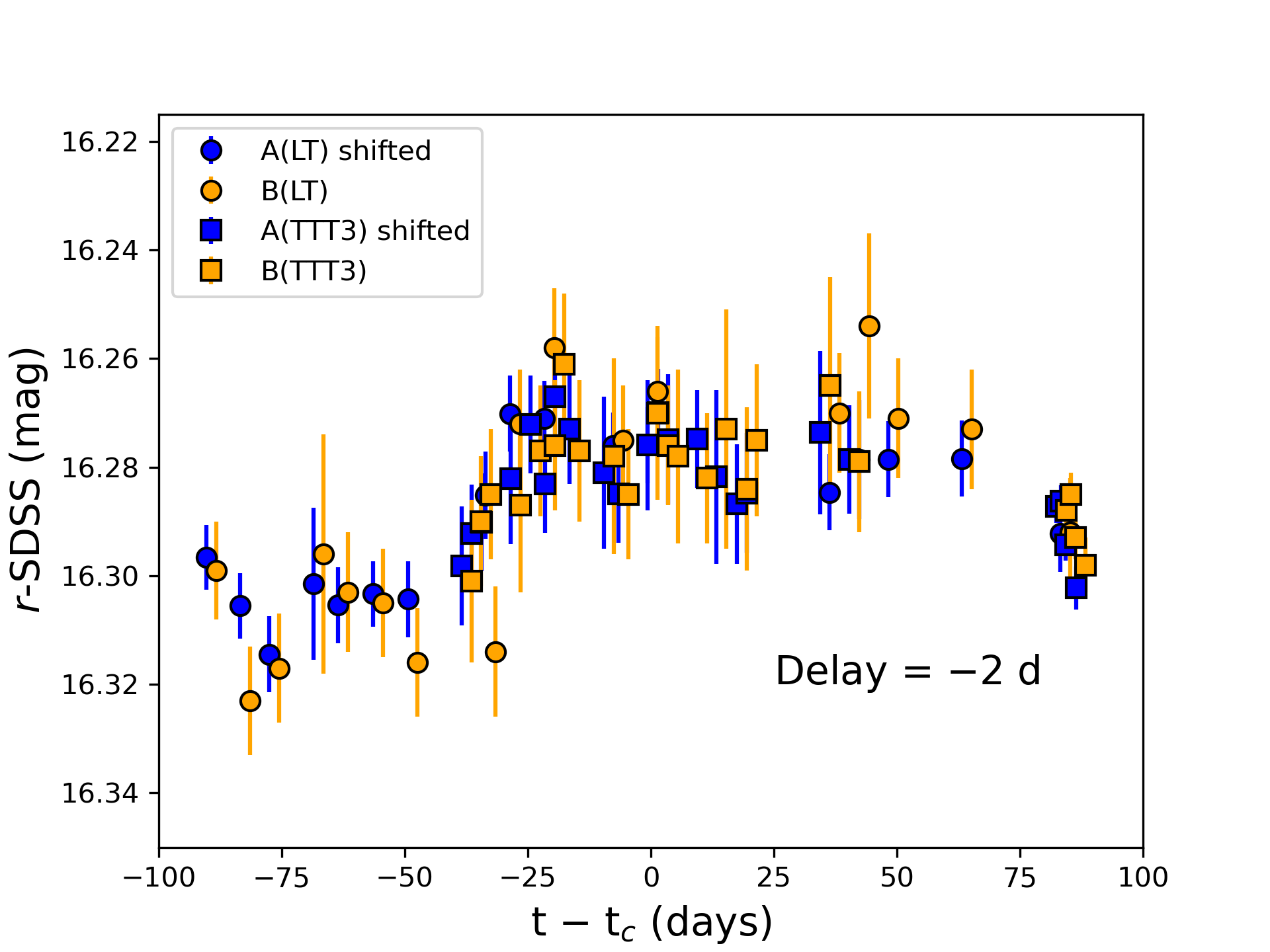}
\includegraphics[width=0.3\textwidth, trim=2 0 35 5, clip]
{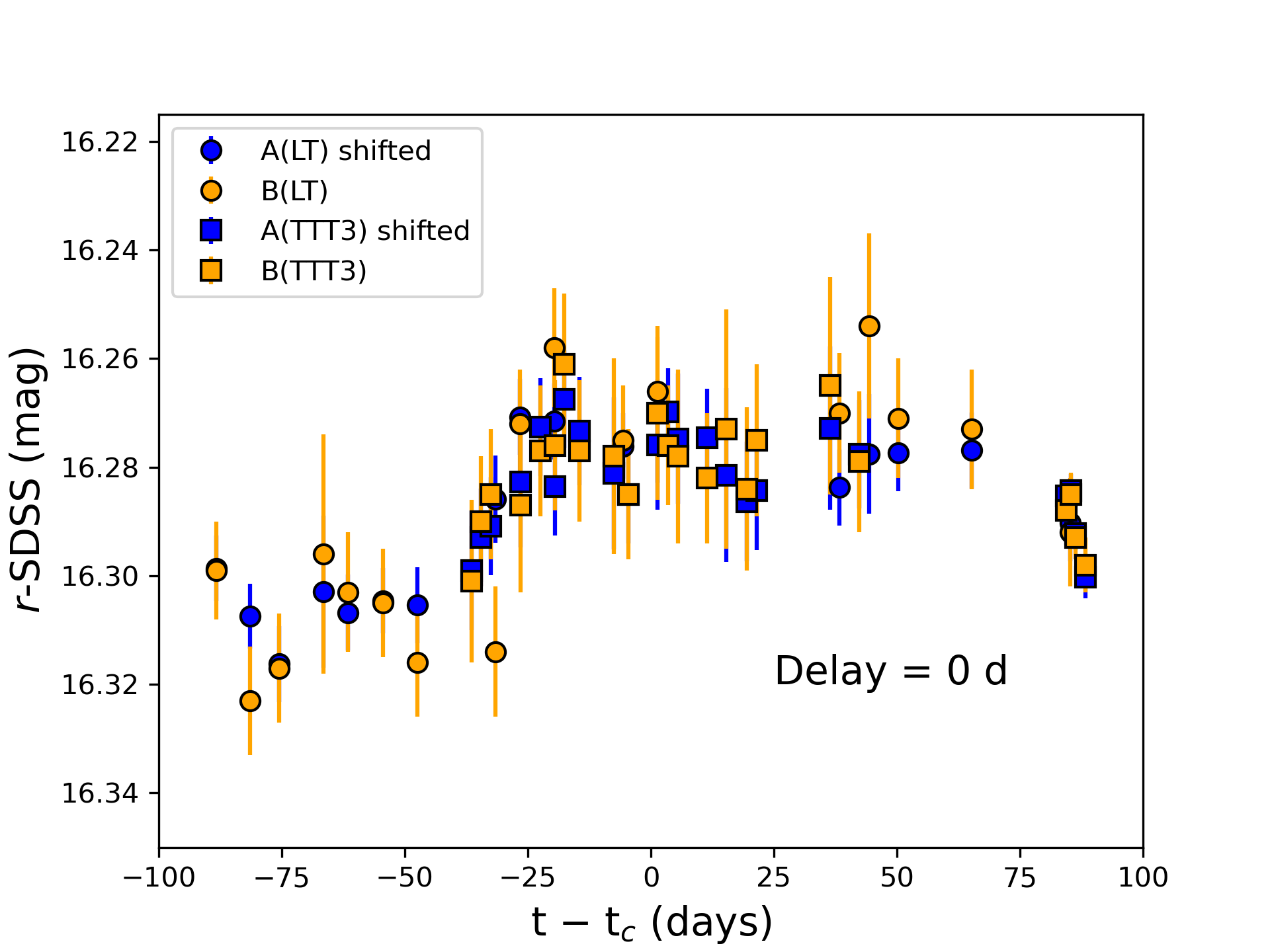}
\includegraphics[width=0.3\textwidth, trim=2 0 35 5, clip]
{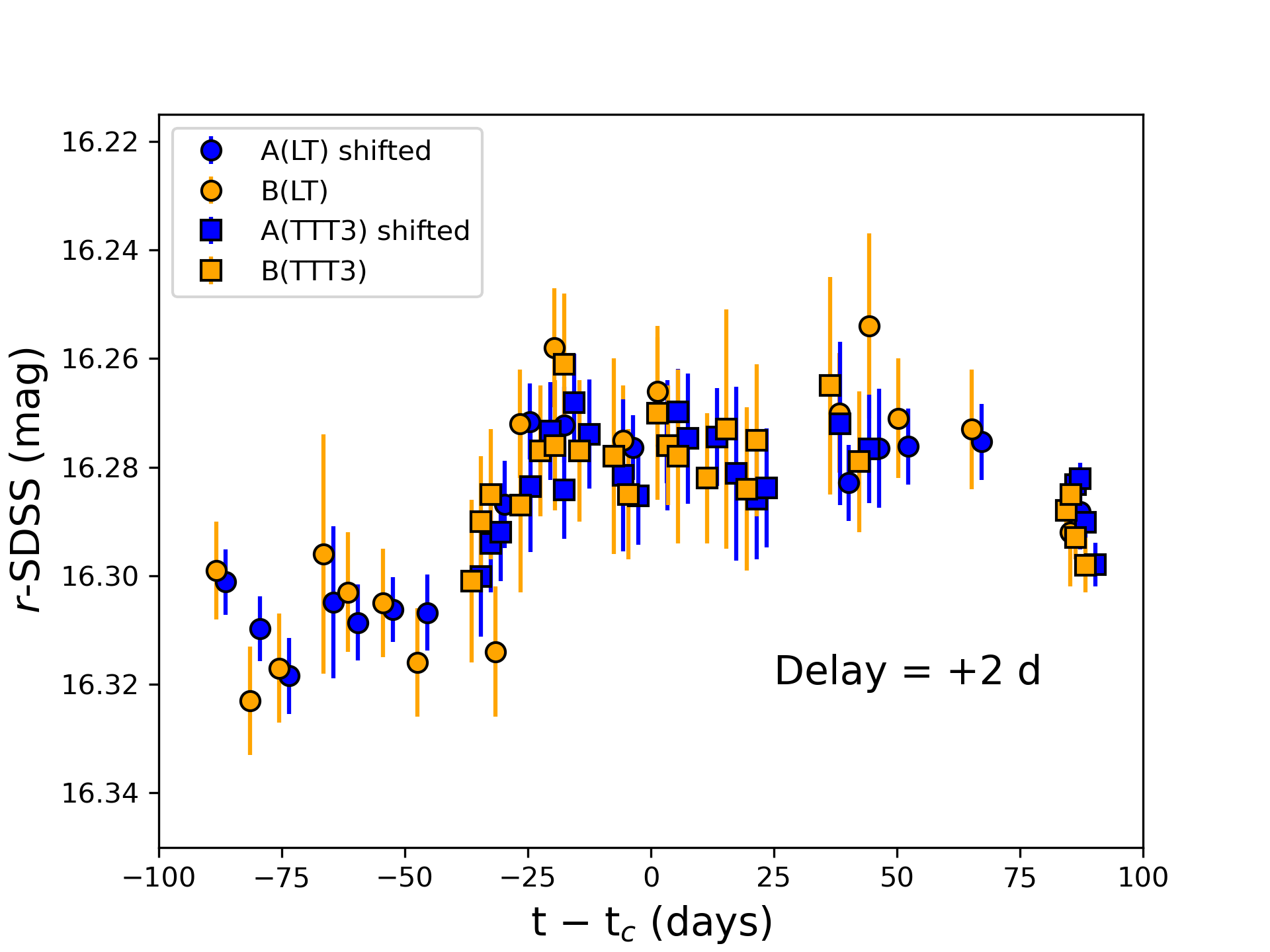}
\includegraphics[width=0.3\textwidth, trim=2 0 35 5, clip]
{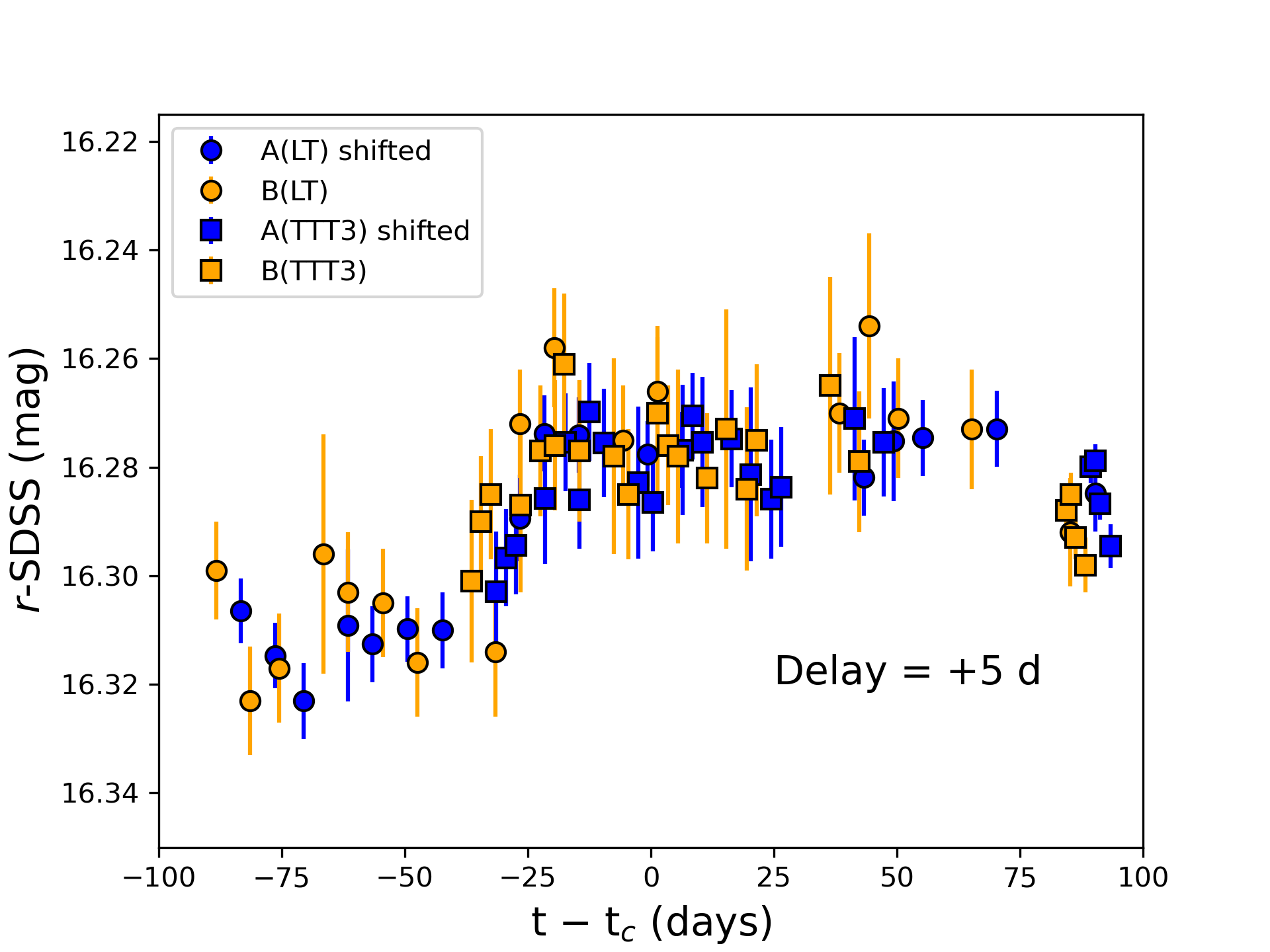}
\caption{Combination of $r$-band light curves of images A and B of \object{PS J0147+4630}. The original curve of B and shifted records of A are drawn together. To shift the curve of A, we consider five delays and the corresponding linear microlensing signals (minimum dispersion for a Gaussian decorrelation length of 4 d). Here, t$_{\rm c}$ represents a central observing epoch on 18 October 2025.}
\label{fig:parach_delayAB}
\end{figure*}

\justifying The histogram in the right panel of Figure~\ref{fig:parach_d2AB} indicates that the $r$-band AB delay is expected to be in the range of ($-$2, $+$2) d, which basically coincides with the 2$\sigma$ confidence interval in Table~\ref{tab:ABCdelay}. This delay histogram also rules out delays of $\pm$5 d. To test the reliability of these results, in Figure~\ref{fig:parach_delayAB} we combine the light curve of B and the brightness record of A, shifted in magnitude and time. To create the five combined records, we use delays of $-$5, $-$2, 0, $+$2, and $+$5 d, and microlensing signals (linear magnitude offsets) that minimise dispersion for those delays. It is worth noting that the measurement of a near-zero delay was performed using light curves containing a significant intrinsic signal, since the signal-to-noise ratio is $\sim$4$-$5. Additionally, the shape of the intrinsic variations also plays a role, and thus, the rise in brightness around day $-$35 and the drop from day 75 allow us to visually verify that, as expected, there is a clear discrepancy between the original curve of B and the shifted curve of A for delays of $\pm$5 d.

\FloatBarrier

\section{Posterior distributions from LENSTRONOMY MCMC results for \object{PS J0147+4630}}
\label{sec:lenstronomyplot} 
\centering
\vspace{0.5cm}
\includegraphics[width=0.85\textwidth,
trim=5 5 5 0, clip]{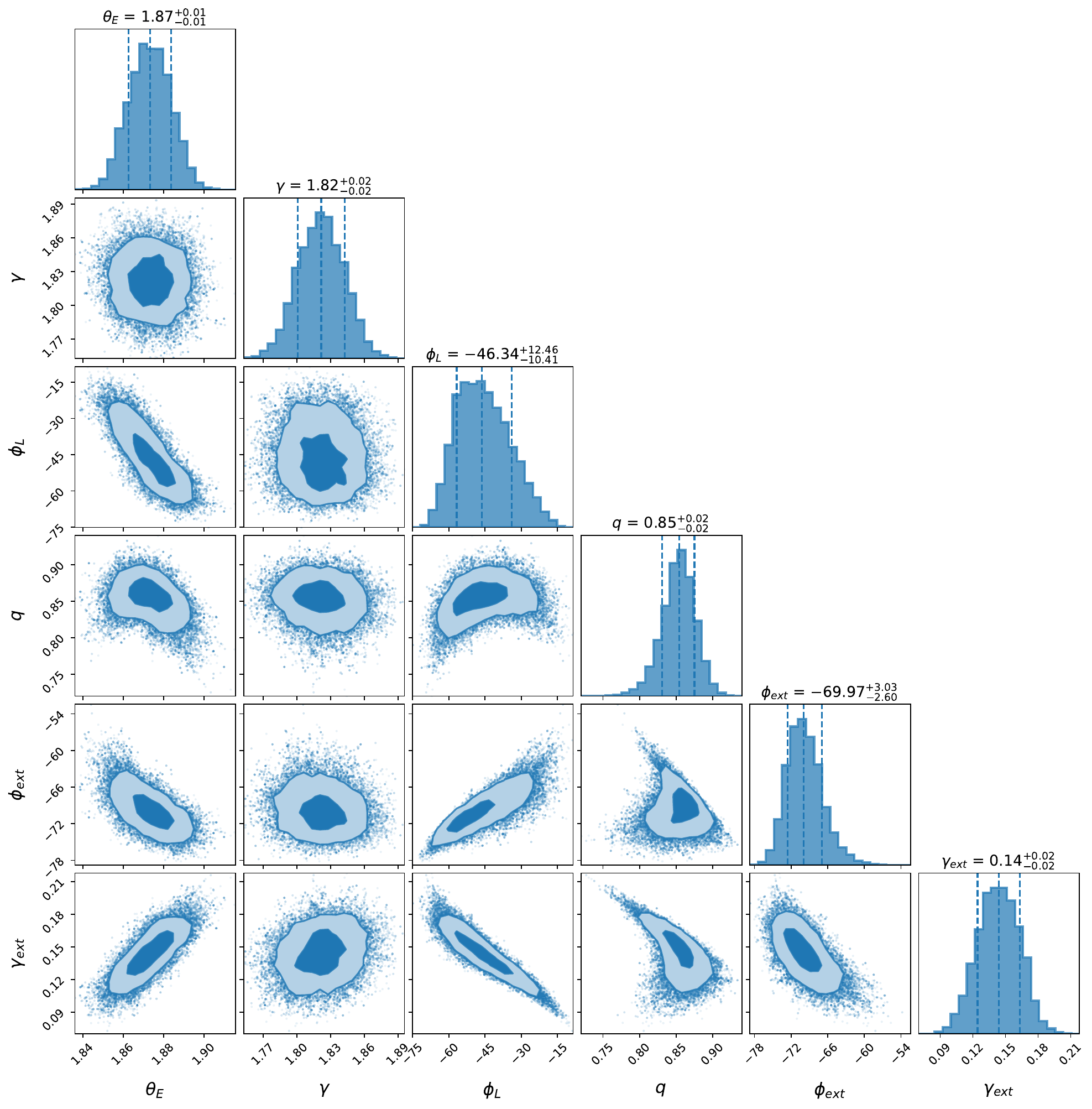}
\caption{Posterior distributions for the parameters of the SPLE+ES model using the LENSTRONOMY software with HST astrometry constraints for the quad \object{PS J0147+4630} and a fiducial cosmology with $h=0.7$. The contour lines correspond to the 1-$\sigma$ and 2-$\sigma$ limits and the dashed lines mark the median $\pm 1\sigma$ levels. Position angle of the lens ($\phi_L$) and external shear angle ($\phi_{ext}$) are given in degrees from E to N.}
\label{fig:cornerplot}
\FloatBarrier

\end{appendix}

\end{document}